\documentclass[a4paper, 11pt]{article}

\usepackage[hang,flushmargin]{footmisc}
\usepackage{siunitx}
\usepackage{jinstpub}
\usepackage{array}
\newcolumntype{x}[1]{>{\centering\arraybackslash\hspace{0pt}}p{#1}}
\DeclareSIUnit{\E}{E}

\newcommand{\w}{w}
\newcommand{\pitch}{d}
\newcommand{\thicc}{t}

\title{Effect of Diffusion on the Peak Value of Energy Loss Observed in a LArTPC}
\author[a]{G. Putnam,}
\author[a]{D.\,W.~Schmitz.}
\affiliation[a]{Enrico Fermi Institute, University of Chicago, Chicago, IL, 60637, USA}
\emailAdd{grayputnam@uchicago.edu}

\usepackage[x11names,dvipsnames,table]{xcolor} 
\usepackage{colortbl}
\usepackage{listings}
\abstract{
Liquid Argon Time Projection Chamber (LArTPC) detectors observe ionization electrons to measure
charged particle trajectories and energy. In a LArTPC, the long time ($\sim$ms) between when the ionization is produced and
when it is collected
means that diffusion can smear the charge by an amount comparable to the spatial
resolution of the detector, given by the spacing between charge sensing channels ($\sim$mm).
This smearing has an
impact on the distribution of energy losses measured by each channel. In particular,
the smearing increases the length of the charged particle trajectory observed by each channel, and therefore the
most-probable-value (MPV) of particle energy loss recorded by that channel. We find, for example, that this effect
shifts the MPV $dE/dx$ of a muon with an energy of \SI{1}{GeV} by ${\sim}4$\% for a \SI{2}{ms} drift time and
\SI{4.7}{mm} wire spacing, as
in the DUNE-FD LArTPC.
This has implications for the energy-scale
calibration and electron lifetime measurements in a LArTPC, which both use the MPV of the muon
energy loss distribution as a ``standard candle''. The impact of diffusion on these calibrations is
assessed.}

\newenvironment{linenomath}
{
}%
{}

\arxivnumber{2205.06745}

\begin{document}
\maketitle
\flushbottom

\section{Introduction}

Liquid Argon Time Projection Chamber (LArTPC) detectors primarily leverage the ionization charge
produced by charged particles traversing the medium to reconstruct their energy loss profile
for use in particle identification and
kinematic reconstruction. In a LArTPC, charge is drifted from the particle ionization column to a
set of readout channels by the effect of a large electric field. The readout channels can be a plane
of wires \cite{Carlo,ICARUSDet,ArgoNeuTD,MicroBooNEDet,SBNDet,ProtoDUNEDet,DUNEDet} or an array of pixels \cite{pix} depending on the detector.
Between production and collection a number of effects
perturbs the charge: impurities in the argon absorb electrons, distortions to the electric field
(such as those from space-charge \cite{uBSpaceCharge}) deflect the ionization from its path, and diffusion smears out
the charge. These effects can cause non-uniformities in the detector response, and typically the
calibration procedure in a LArTPC experiment will analyze and eliminate them to make the detector response uniform.
This work focuses on the role of diffusion between ionization production and collection: we find that
diffusion has a significant impact on the probability distribution of charge collected by each
channel which has not been appreciated in previous LArTPC experiments. In contrast to other
effects,
which are treated as perturbations to the detector response, the role of diffusion should be
understood as changing the underlying distribution of energy loss recorded by each channel.

The probability distribution governing particle energy loss observed by a
channel is the Landau-Vavilov distribution \cite{Vavilov:1957zz}. In the limit that the charged
particle is relativistic and the slice of the particle track that the channel is sensitive to 
is small (as is often the case in a LArTPC), this distribution is well-approximated by the Landau
distribution \cite{Landau:1944if}. 
For both distributions, the mean energy loss per length (given by the Bethe-Bloch theory
\cite{Bethe})
is independent of the length of the charged particle trajectory observed by the
channel. However, the shape of the distribution depends strongly on
this length. 

The Landau and Landau-Vavilov distributions are governed by the
cross-section of charged particles incident on atomic electrons, which has a power law
dependency on the energy transfer. This behavior means that
the mean energy loss is influenced by a small number of large energy transfers to electrons
well above the atomic excitation energy
($\delta$-rays). When the channel is only sensitive to a short enough length of
the charged particle trajectory such that it will not observe a $\delta$-ray most of the time,
the bulk of the distribution is below the mean value of energy loss with a long tail
extending out to high energy losses. 
The Landau distribution is an
appropriate approximation in this case. 
As the channel-sensitive length
increases, more of the delta
rays get absorbed into the bulk of the distribution and the location of the peak shifts up as the variance
drops, until the distribution is better modeled by a Landau-Vavilov distribution. 

Traditionally in LArTPC experiments, the length of the charged
particle track observed by each channel has been
understood as being given by the spacing between channels.
However, diffusion changes this length. Between production and collection, ionization electrons in a LArTPC diffuse in the direction of their drift
velocity (``longitudinal'' diffusion) and in the two perpendicular directions (``transverse''
diffusion). The magnitude of the smearing is different between the longitudinal and transverse
directions and is parameterized by two separate diffusion constants. In the transverse
directions, the smearing is given by \cite{Einstein,Smoluchowski}
\begin{equation}
	\sigma_\text{T} = \sqrt{2 \cdot D_\text{T} \cdot t_\text{drift}},
	\label{eq:smearwidth}
\end{equation}
where $D_\text{T}$ is the transverse diffusion constant and
$t_\text{drift}$ is the time between production and collection for a
cluster of ionization charge. 
The transverse diffusion constant has never been directly measured in LAr at the electric fields
typical in existing LArTPC neutrino detectors (although it has been measured at higher fields
\cite{DiffT, DiffT2}).

It can in principle be extrapolated from
measurements of the longitudinal diffusion constant $D_\text{L}$ \cite{Wannier,Wannier2}, which
have been performed \cite{uBDL,LiDL,ICARUSDL}. 
However, existing longitudinal diffusion measurements are in tension with each other, so there is not a
clear way to perform this extrapolation. From this picture there are two available predictions
for the transverse diffusion constant: one from the Li et al. measurement \cite{LiDL} and
another from 
Atrazhev-Timoshkin theory \cite{Atrazhev}.\footnote{
\noindent Reference \cite{uBDL} points out that to apply Atrazhev-Timoshkin theory, one needs to interpolate from
lower fields to get a prediction at LArTPC electric field values; we use their interpolation for
this computation. Both predictions are specifically for the longitudinal electron energy $\epsilon_{L}$, which
relates to the diffusion constant as $D_{L} = \mu\cdot \epsilon_{L} / e$ 
\cite{Einstein,Smoluchowski}, where $\mu$ is the
electron mobility and $e$ is the electric charge. 
As detailed in \cite{LiDL},
$D_L$ then relates to $D_T$ through the generalized Einstein relations first proposed by
Wannier \cite{Wannier,Wannier2} (which also depend on $\mu$).
To get the transverse diffusion constant, we apply the
electron mobility parameterization from the Li et al. work \cite{LiDL}.
}
At a drift field of \SI{500}{V\per cm} (typical for existing LArTPCs), 
the Li et al. (Atrazhev-Timoshkin) parameterization predicts
a transverse diffusion constant of \SI{12.0}{cm\squared\per\second}
(\SI{9.30}{cm\squared\per\second}) and that the transverse diffusion width
in time is
\begin{equation}
	\sigma_\text{T} = 1.55 \ (1.36) \ \sqrt{\frac{t_\text{drift}}{1\text{ms}}} \ \text{mm}
	\,.
\end{equation}

\begin{figure}[t]
	\centering
	\includegraphics[width=0.45\textwidth]{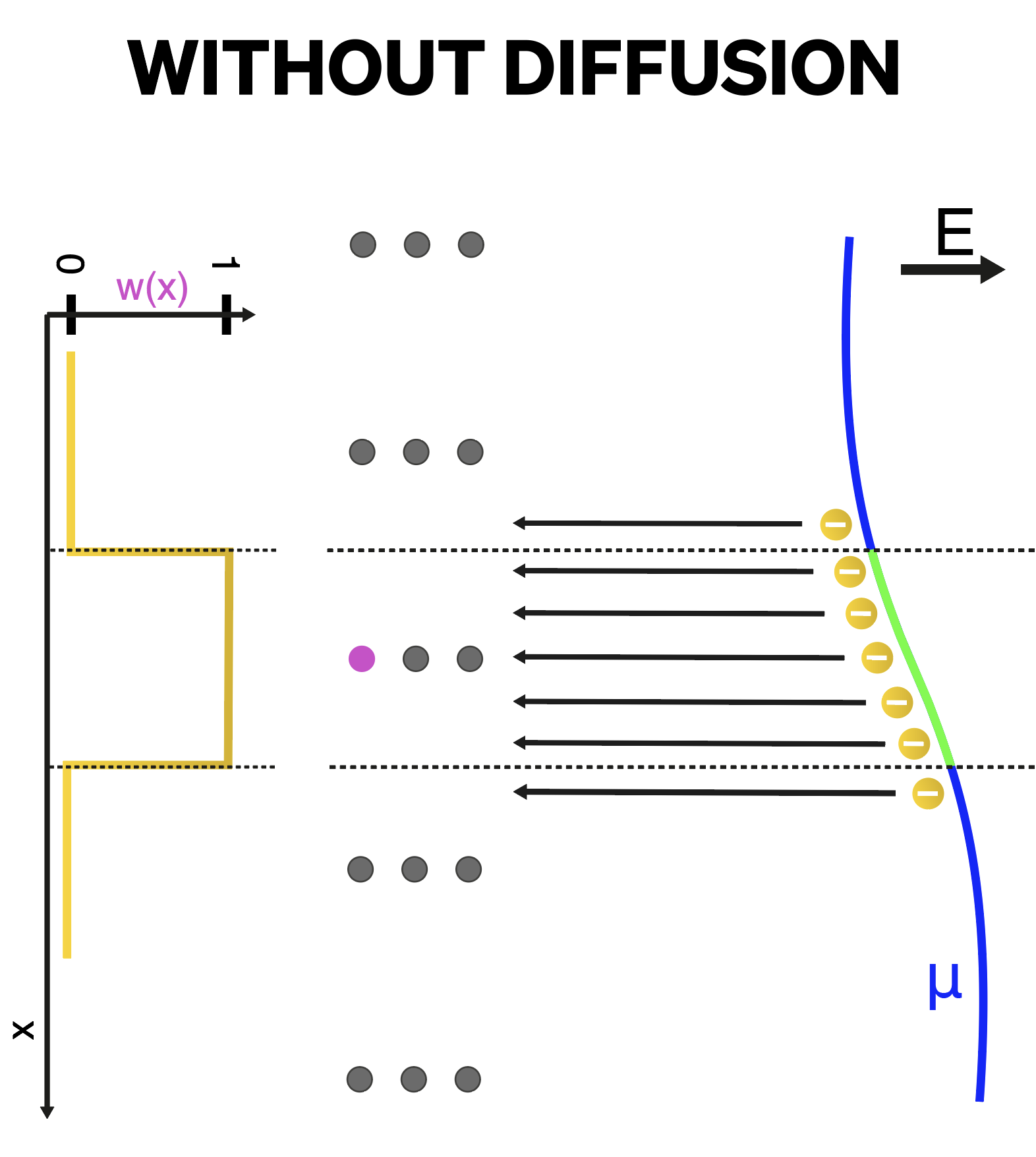}
	\hspace{1cm}
	\includegraphics[width=0.45\textwidth]{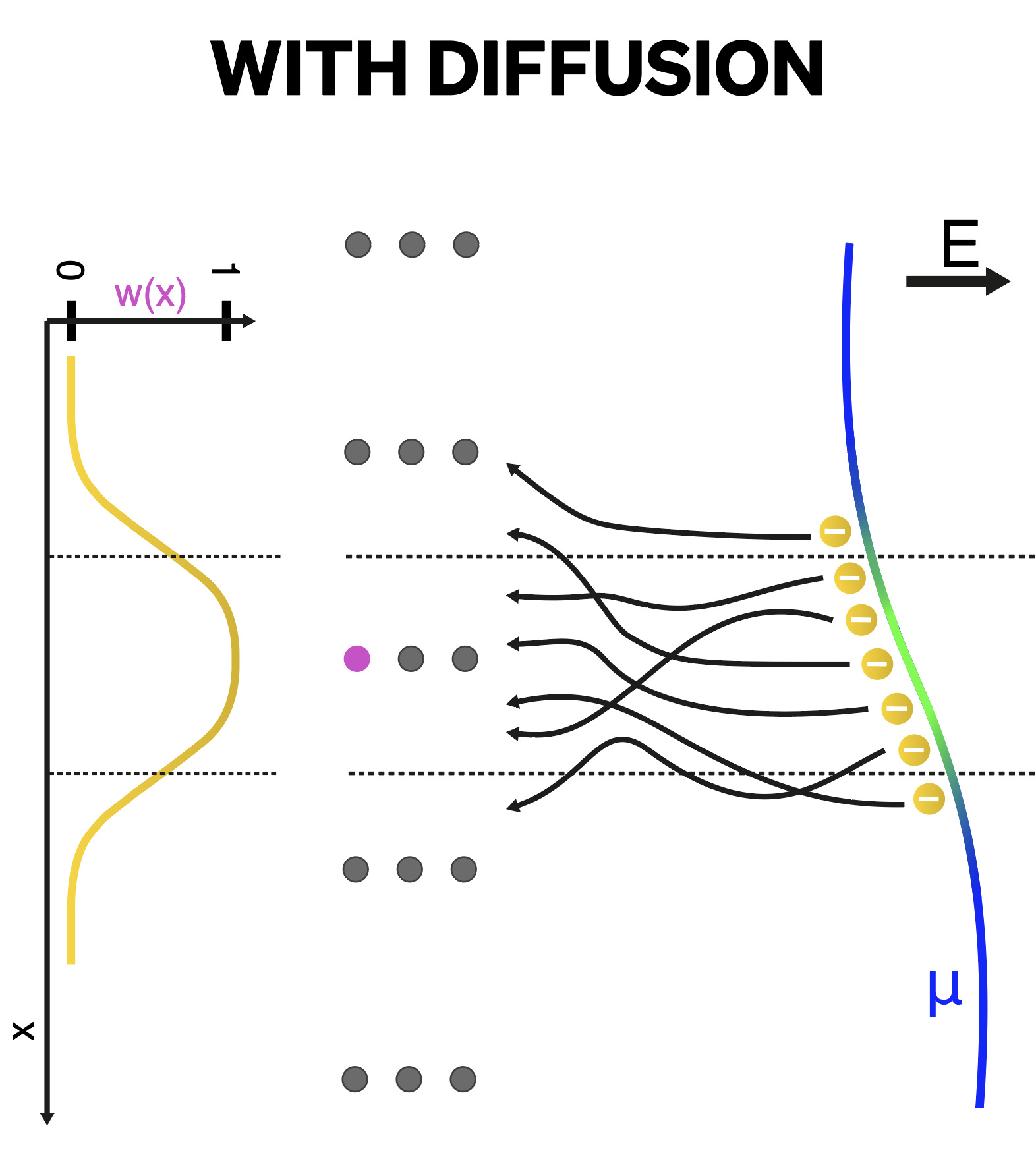}
	\caption{Diagrams illustrating the effect of diffusion on the path length of a particle track
	observed by
	channels in a LArTPC with a wire plane readout. The graphs on the left of both diagrams
	plot the weight function, $w(x)$, describing the probability that ionization energy is
	observed as observed by the central, pink wire as a function of
	the position of the energy deposition along the particle track (this is also denoted by
	the color of the particle track).
	Diffusion in the direction transverse to the wire orientation smears charge in and
	out of the range each wire is sensitive to.}
	\label{fig:diffusion}
\end{figure}

Diffusion in the direction perpendicular to the drift
orientation smears the positions of ionization electrons observed by each channel.
Figure \ref{fig:diffusion} illustrates the case of a plane of wires in a LArTPC.
The diffusion in the perpendicular direction means that
each channel sees a larger
length of the charged particle track than would just be given by the channel spacing: \emph{diffusion
effectively makes the spacing a drift-dependent quantity}. This means that
the Landau-Vavilov or Landau distribution of energy loss observed by each channel is influenced by
diffusion. In particular, diffusion acts to increase the most-probable-value (MPV) of energy
loss recorded by each channel.

This effect is derived and discussed from a
generalized perspective in section
\ref{sec:dist}, with further details in appendix \ref{sec:derivation}.
Then, in section \ref{sec:wlartpc}, we apply the general results of
section \ref{sec:dist} to the LArTPC case, and in section \ref{sec:calib} we discuss the outlook for
LArTPC calibrations.

\section{Distribution of Energy Loss Seen by a Generalized Channel}
\label{sec:dist}

The Landau-Vavilov and Landau distributions are derived assuming that a detector 
is sensitive to a discrete section of the charged particle trajectory:
there is a region where the channel detects all
of the deposited energy, and a region where it detects none. In the case of a LArTPC impacted by
diffusion, this assumption does not hold. As is illustrated in figure \ref{fig:diffusion}, a
channel
in a LArTPC is sensitive to a fractional amount of the particle energy loss, dependent on its
position. To parameterize this effect, we define the channel ionization weight function $\w(x)$, 
which gives this
fraction as a function of the particle position $x$. This weight function defines the
probability distribution of energy that the channel measures. 

To obtain the distribution of energy loss for a general channel ionization weight function, we go through the
usual derivation of the Landau-Vavilov distribution, keeping this weight function along the way.
Section \ref{subsec:principles} introduces the principles in this derivation. Section
\ref{subsec:analytic} gives
an analytic formula for the energy loss distribution. 
Section \ref{subsec:landau} discusses the
relativistic/thin-film limit for a weight function. Details of the derivation are given in appendix
\ref{sec:derivation}. Table \ref{tbl:symbols} lists the important symbols in the derivation.

\begin{table}[htbp]
\centering
\rowcolors{2}{gray!25}{white}
\begin{tabular}{x{2cm} | x{9cm} | x{3cm}}
	\rowcolor{gray!15}
	Symbol & Definition & Units (length L, time T, energy E)\\
	\hline
	$\sigma_T$ & Transverse diffusion width (equation \ref{eq:smearwidth}) & L\\
	$D_T$ & Transverse diffusion constant (equation \ref{eq:smearwidth}) &
	\si{L \squared\per T}\\
	$\frac{d\sigma}{dT}$ & Differential cross section of particle incident on (bare) electrons 
	w.r.t.\ energy transfer (equation
	\ref{eq:xsec}) & \si{L\squared\per E}\\
	$\zeta$ & Strength of scattering on bare electrons (equation \ref{eq:zeta}) &
	\si{\per\L}\\
	$w(x)$ & Channel ionization weight function: weight given to energy deposited at a
	position $x$ as measured by a channel (defined for wire plane LArTPCs in equation
	\ref{eq:wlartpc}) & --\\
	$p_\w(E)$ & Probability distribution of energy ($E$) measured by a channel with a weight
	function $w$ (general case in equation \ref{eq:LandauVavilov} and in the Landau limit in
	equation \ref{eq:landaulimit}) & \si{\per\E}\\
	$\pitch$ & Channel pitch: the spacing between charge sensors (equation \ref{eq:pitch}) & \si{\L}\\
	$\thicc$ & Channel thickness: the length of a particle track observed by a charge
	sensor, distinct from $\pitch$ in the presence of diffusion (equation \ref{eq:thicc}) & \si{\L}\\
\end{tabular}
\caption{Important symbols referenced throughout the text.}
\label{tbl:symbols}
\end{table}

\subsection{Principles of the Energy Loss Distribution}
\label{subsec:principles}

We construct the probability distribution of energy loss iteratively. We start from the
simple base case of the distribution over an infinitesimal length, and then build it up into the
distribution over finite length. Our tool in this construction is the convolution property:
given the probability
distribution of observing a particle energy loss $E$ over a length $\ell$, $p_\ell(E)$, a
convolution of $p_\ell(E)$ with itself doubles the length $\ell$ since a convolution represents all the
ways a particle can lose an amount of energy in two steps of $\ell$.
Thus the convolution property states that
\begin{equation}
	p_{2\ell}(E) = \int p_\ell(E-E')p_\ell(E') dE' \,.
	\label{eq:conv}
\end{equation}

The distribution of energy loss for an infinitesimal length is given by
the limit that the particle travels a short enough distance such that it will scatter off at
most one atomic electron. In this case, the probability of energy loss is equal to the probability of
colliding with an atomic electron with precisely that amount of energy transfer.
The probability of no energy loss is equal to the probability of not
interacting with an atomic electron. Thus,
\begin{equation}
	\underset{\ell\to 0}{\mathrm{lim}} \ p_\ell(T) = (1 - \rho \sigma \ell) \delta(T) + \rho\ell
	\frac{d\sigma}{dT}
	\label{eq:smalll}
	\,,
\end{equation}
where $\rho$ is the number density of electrons, $\frac{d\sigma}{dT}$ is
the differential cross section w.r.t.\ the energy transfer $T$, $\sigma = \int
\frac{d\sigma}{dT} dT$, and $\delta$ is the Dirac-Delta function.

In the case of heavy charged particles (such as muons and protons) elastically
scattering on bare electrons, the cross section is given by the
Rutherford formula \cite{PDG}
\begin{equation}
	\frac{d\sigma}{dT} = \frac{2\pi r_e^2 m_e}{\beta^2}\frac{1-\beta^2 T /
	T_\text{max}}{T^2}
	\label{eq:xsec}
	\,,
\end{equation}
where $r_e$ is the classical electron radius, $m_e$ is the electron mass, $\beta$ is given by
the particle velocity, and $T_\text{max}$ is the maximum energy transfer in a single collision,
\begin{equation*}
T_\text{max} = \frac{2 m_e \beta^2 \gamma^2}{1 + 2\gamma m_e/M + (m_e/M)^2}
\,,
\end{equation*}
where $M$ is the mass of the charged particle. The cross section for atomic electrons is
modified relative to the bare cross section at low energy transfer by atomic effects.
A useful quantity related to the cross section is
\begin{equation}
	\label{eq:zeta}
	\zeta = \rho \frac{2\pi r_e^2m_e}{T_\text{max} \beta^2}
	\,.
\end{equation}
$\zeta$ is a quantity with units of inverse length that encodes the 
rate of scattering.

For the probability distribution of energy loss over an infinitesimal length, the channel
ionization weight function
can always be assumed to be a constant $\w$. In this case, 
for a channel to measure an energy deposition $E$ the particle must deposit an
amount $E/\w$. So in the presence of a weight function $\w$, $p_\ell(E)$ transforms to
$\frac{1}{\w}p_\ell(E/\w)$ 
(the $1/\w$ in front fixes the normalization) when the length $\ell$ is small.

\subsection{Analytic Form of the Distribution}
\label{subsec:analytic}
To build the particle energy loss distribution, we discretize the weight function $\w(x)$ into weights $\w_i$ over infinitesimal steps
$dx_i$, and build up the probability distribution over the full weight function by performing a
product of convolutions:
\begin{equation}
	p_\w(E) = \int d T_1 \int d T_2 \cdots \int d T_n \frac{p_{dx_0}(\frac{E - T_n - \cdots -
	T_1}{\w_0})}{\w_0}
	\times
	\frac{p_{dx_1}(\frac{T_n - \cdots - T_1}{\w_1})}{\w_1} 
	\times \cdots \times
	\frac{p_{dx_n}(\frac{T_n}{\w_n})}{
	\w_n}
	\,.
\end{equation}
Appendix \ref{subsec:derivation} goes through the process of simplifying this into an algebraic
expression for the case of Rutherford scattering. In the process of this derivation, it is
necessary to introduce the mean energy loss $\overline{E}$ as an external input to the
distribution. This is because atomic effects modify the bare electron cross section at low
energy transfer
in a way that can be accounted for through the mean energy loss using Bethe-Bloch theory. 
We obtain:
\begin{equation}
\begin{split}
p_\w(E) = 
\frac{1}{2\pi\zeta T_\text{max}}
\int\limits_{-\infty}^{\infty} dz \ \mathrm{exp}
\left[\frac{iz}{\zeta T_\text{max}}
\vphantom{\int} \right. &
(E-\overline{E}) 
+ \int dx  \ \zeta ( 1
 - e^{-i \w(x) z/\zeta}) 
 - i z \w(x) (1 + \beta^2) + \\
& (\zeta\beta^2 + i \w(x) z)(-\text{Ei}[-i \w(x) z/\zeta] +
\text{log}[ i \w(x) z/\zeta] + \gamma_\text{EM})
\left. \vphantom{\int} \right] 
\label{eq:LandauVavilov}
\,.
\end{split}
\end{equation}
where $\gamma_\text{EM} \approx 0.577$ is the Euler-Mascheroni constant
and $\text{Ei}$ is the exponential integral function,
$\text{Ei}(x) = -\int\limits_{-x}^\infty dt \ e^{-t}/t$.
This is the probability distribution of particle energy loss for a channel with a general
ionization weight function $\w(x)$ as a function of the charged particle position.
Appendix
\ref{subsec:vavilov} shows that this distribution reduces to a Landau-Vavilov distribution precisely when the
channel
ionization weight function is given by a step function.

\subsection{The Thin-Film/Relativistic Limit}
\label{subsec:landau}

To take the thin-film/relativistic limit, we take the 
case where the weight function is narrow compared to the size of
the scattering length $1/\zeta$ (details in appendix \ref{subsec:landauapp}). In this limit,
the probability distribution of energy loss converges to
\begin{equation}
        p_\w(E) = \frac{1}{2\pi i \zeta T_\text{max} \pitch}
        \int\limits_{-i\infty}^{i\infty} dz' \ \text{exp} \left[
        z' \left(
	\lambda
        + \text{log}|z'|\right)\right]
	\label{eq:landaulimit}
        \,,
\end{equation}
where
$\lambda = \frac{E - \overline{E}}{\zeta
T_\text{max}\pitch} - \text{log}\zeta \pitch
+ \gamma_\text{EM} - 1 - \beta^2 + \frac{\int dx \ \w(x) \text{log}[\w(x)]}{\pitch}$ and
$\pitch$ is defined as the pitch:
\begin{equation}
	\pitch\equiv \int \w(x)dx
	\label{eq:pitch}\,.
\end{equation}
Equation \ref{eq:landaulimit} can be recognized as the Landau distribution for a parameter $\lambda$,
which has a maximum at
$\lambda_\text{MPV}
= \gamma_\text{EM} - 0.8 \approx -0.22278\ldots$\cite{MPVval}. Therefore:
\begin{linenomath}\begin{equation*}
        E_\text{MPV} = 
	\overline{E} + \zeta T_\text{max} \pitch \left( \text{log}\zeta\pitch + 0.2
        + \beta^2 - \frac{\int dx \ \w(x) \text{log}[\w(x)]}{\pitch}\right)
        \,.
\end{equation*}\end{linenomath}
In the case where the weight function $\w(x)$ is a step function, $E_\text{MPV}$ reduces to the usual result \cite{PDG}.
We can recast this result by defining the thickness
\begin{equation}
	\thicc \equiv \pitch \hspace{0.05em} e^{-\int dx \ \w(x) \text{log}[\w(x)]/\pitch}
        \label{eq:thicc}
        \,.
\end{equation}
Then the formula for the MPV energy loss $dE/dx_\text{MPV} \equiv
E_\text{MPV}/\pitch$ is
\begin{equation}
                \frac{dE}{dx}_\text{MPV} 
                = \overline{\frac{dE}{dx}} + \zeta T_\text{max} \left( \text{log}\zeta\thicc + 0.2
        + \beta^2\right)
	\label{eq:landauMPV}
        \,. 
\end{equation}
This equation, with equation \ref{eq:thicc} as an input, is the main result of this derivation.
In this formula, all the dependence on the weight function $\w(x)$ is encoded in the
$\thicc$ parameter. This is precisely the usual formula for the energy loss MPV $dE/dx$,
just with the thickness given by equation \ref{eq:thicc}. The thickness and pitch transform
under scaling and dilation of the weight function as:
\begin{equation}
\begin{aligned}[l]
	\text{Scaling}\\
	\w(x) \to a\cdot \w(x)\\
	\pitch \to a\cdot \pitch\\
	\thicc \to \thicc
\end{aligned}
\qquad
\begin{aligned}[l]
	\text{Dilation}\\
	\w(x) \to \w(x/a)\\
	\pitch \to a\cdot\pitch\\
	\thicc\to a\cdot\thicc.
\end{aligned}
\label{eq:ttransform}
\end{equation}

\subsection{Derivation Summary}

In the nominal case (a step-function weight function), in the thin-film/relativistic approximation, 
the distribution of particle energy loss recorded by a channel is a Landau distribution.
For a general weight function in the relativistic limit, the shape of the distribution is
unchanged relative to the nominal case: it is still a Landau distribution. However, the MPV of the
distribution shifts in a way parameterized by the channel thickness defined in equation \ref{eq:thicc}.
Outside of the relativistic/thin-film limit, in the nominal case of the weight function 
the distribution of particle energy loss is a Landau-Vavilov distribution. In general, for any 
weight function
outside of the nominal case, this distribution is not equivalent to a
Landau-Vavilov distribution.
The form of the probability distribution of particle energy loss for a general 
weight function
is given by equation \ref{eq:LandauVavilov}.

These results can be understood in terms of the
stability of the input Landau and Landau-Vavilov distributions. The probability distribution for
a 
weight function
$\w(x)$ is given by the sum of probability distributions for an
infinitesimal length $dx$. At each $dx$, $\w(x)$ is a constant and the infinitesimal distribution is a
Landau distribution (in the relativistic limit) or a Landau-Vavilov distribution (in the general case).
Since the Landau distribution is stable, the sum of all these infinitesimal distributions 
must also be a Landau distribution. In the general case, each infinitesimal probability distribution is a
Landau-Vavilov distribution, which is not stable, and their sum will in general converge to a different
probability distribution.

\section{The LArTPC Channel Ionization Weight Function}
\label{sec:wlartpc}

\begin{figure}[t]
	\centering
	\includegraphics[width=0.6\textwidth]{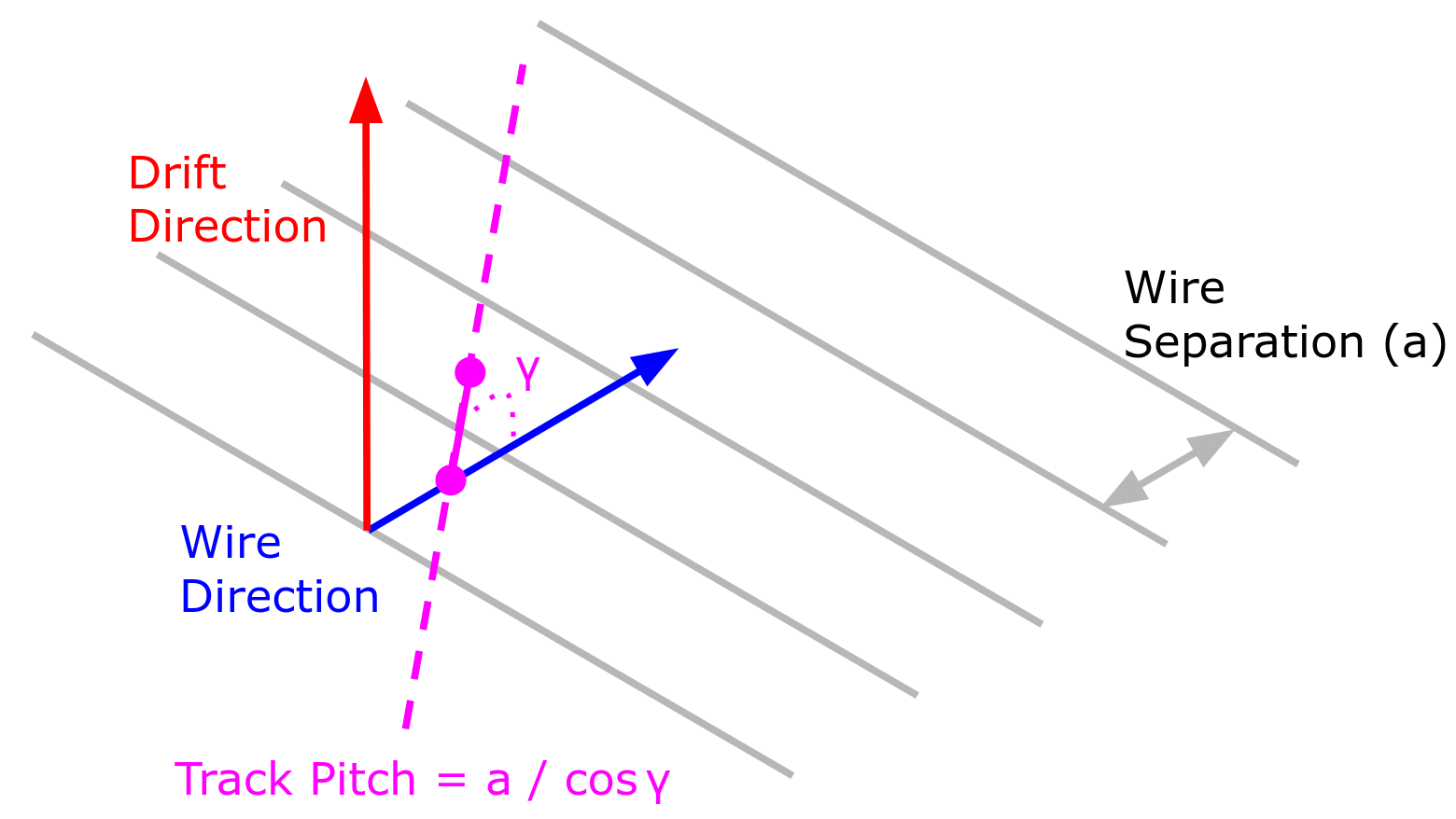}
	\caption{Diagram of the relationship between track orientation and track pitch in a LArTPC.}
	\label{fig:pitch}
\end{figure}

We now apply these results to the case of a LArTPC with a wire plane readout.
LArTPC readout wires detect charge from the induced current of ionization electrons either
passing through the wire plane (a bipolar, ``induction'' signal) or collecting on it (a unipolar,
``collection'' signal). We approximate here that each wire is only sensitive to the ionization
electrons directly adjacent to it, motivating the step function response illustrated in the
left-hand diagram of figure \ref{fig:diffusion} (in the absence of diffusion). 
In reality, electrons can induce a signal as far away as 
\SI{3}{cm} (10 wires for a \SI{3}{mm} spacing) \cite{uBsignal}.
This effect is subdominant, and any change it makes to the channel
thickness will only change the peak energy loss logarithmically (as per equation \ref{eq:landauMPV}). 
Still, assessing the
effect of long-range current induction on the channel ionization weight function merits further study, especially for
induction wire planes. 

To account for the impact of diffusion, the LArTPC channel weight function is modified to be a
convolution of the step function of the wire pitch and a Gaussian smearing induced by transverse
diffusion, as illustrated in the right-hand diagram of figure \ref{fig:diffusion}.
In addition, as
shown in figure \ref{fig:pitch}, a track will in general be at some angle $\gamma$ to the
wire orientation, which dilates the channel weight function $\w$. Putting this together, we obtain
\begin{equation}
	\w_\text{LArTPC}(x) = \int\limits_0^a \frac{dx'}{\sqrt{2\pi}\sigma_\text{T}}
	e^{-(x\cdot\text{cos}\gamma-x')^2/(2\sigma_\text{T}^2)} \,,
	\label{eq:wlartpc}
\end{equation}
for a wire separation $a$ and $\sigma_\text{T}$ given by transverse
diffusion (equation \ref{eq:smearwidth}).

\begin{figure}[t]
	\centering
	\includegraphics[width=0.48\textwidth]{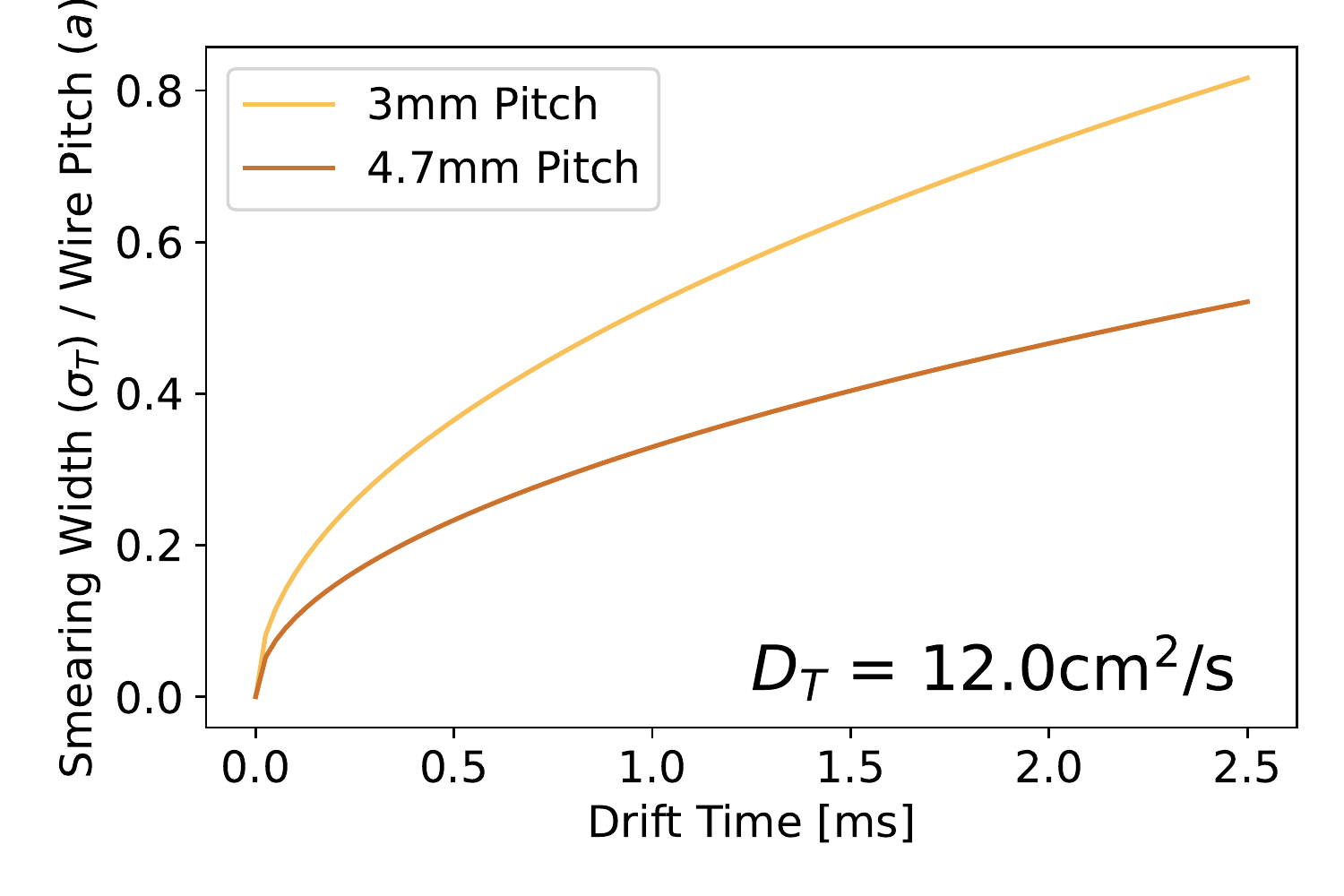}
	\includegraphics[width=0.48\textwidth]{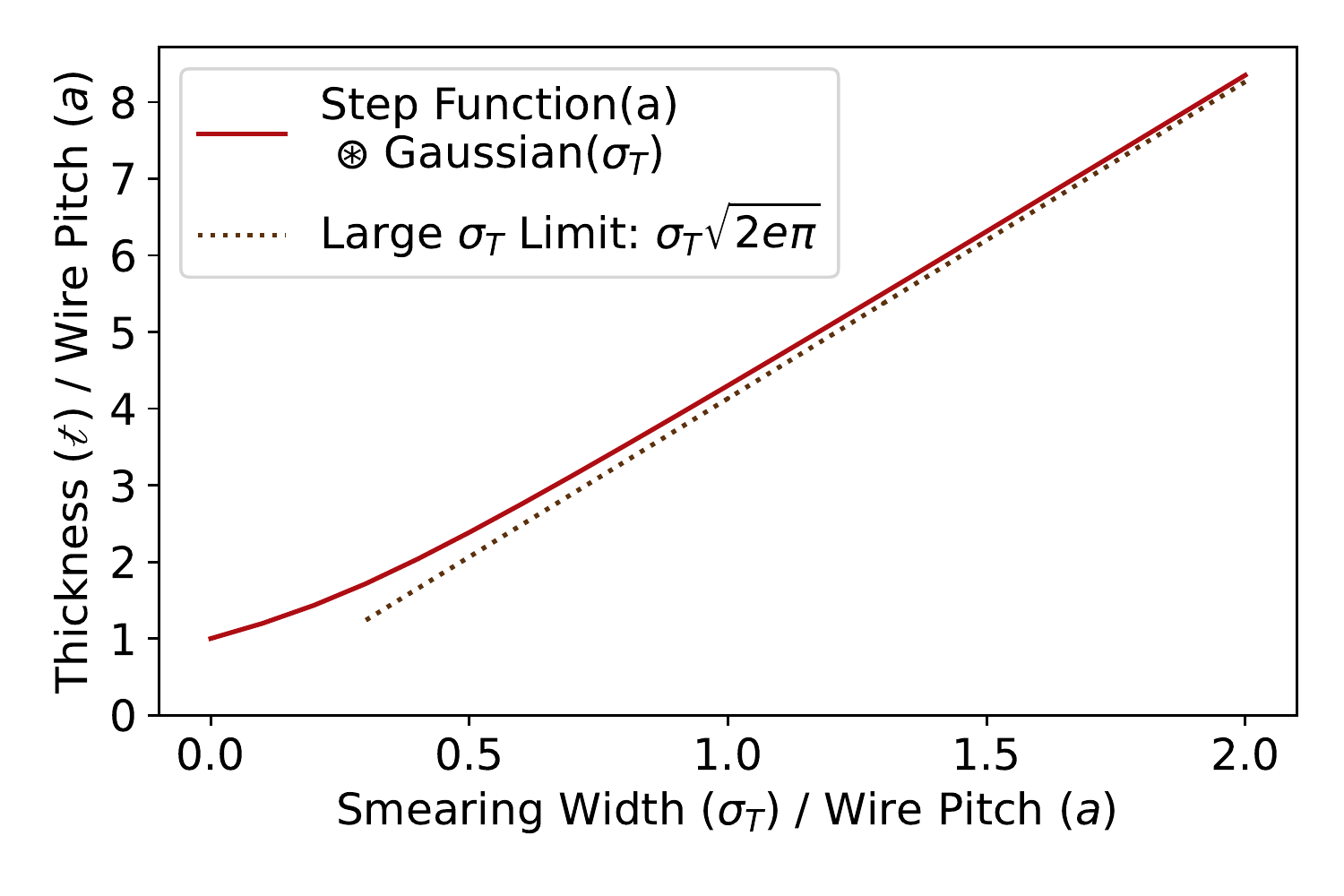}
	\caption{
	(Left) Smearing widths
	from transverse diffusion (using the Li et al. parameterization for $D_T$ \cite{LiDL}) 
	by drift time for the DUNE (\SI{4.7}{mm}) and SBN (\SI{3}{mm}) wire pitches.
	(Right) Thickness observed by a channel with a weight function given by a step function of width $a$ convolved
	with a Gaussian of width $\sigma_\text{T}$, as a function of $\sigma_\text{T} /a$. 
	}
	\label{fig:thicc}
\end{figure}

This weight function has a pitch $\pitch = a/\text{cos}\gamma$ (note that the formula for the
pitch $\pitch$ here coincides with the usual definition of the track pitch in a LArTPC). 
In the limit of a small wire separation relative to the Gaussian width, 
the thickness converges to $\sigma_\text{T}\sqrt{2\pi e}/\text{cos}\gamma$. 
In general though, we have not found a useful closed form of the thickness for this weight
function. A plot
of the thickness $\thicc$ as a function of $\sigma_\text{T} / a$ for $\text{cos}\gamma = 1$ is shown in 
figure \ref{fig:thicc}. By the scaling property of the
thickness under dilation (equation \ref{eq:ttransform}), $\thicc(\text{cos}\gamma) = \thicc(1)/\text{cos}\gamma$. 

\begin{figure}[t]
	\centering
	\includegraphics[width=0.325\textwidth]{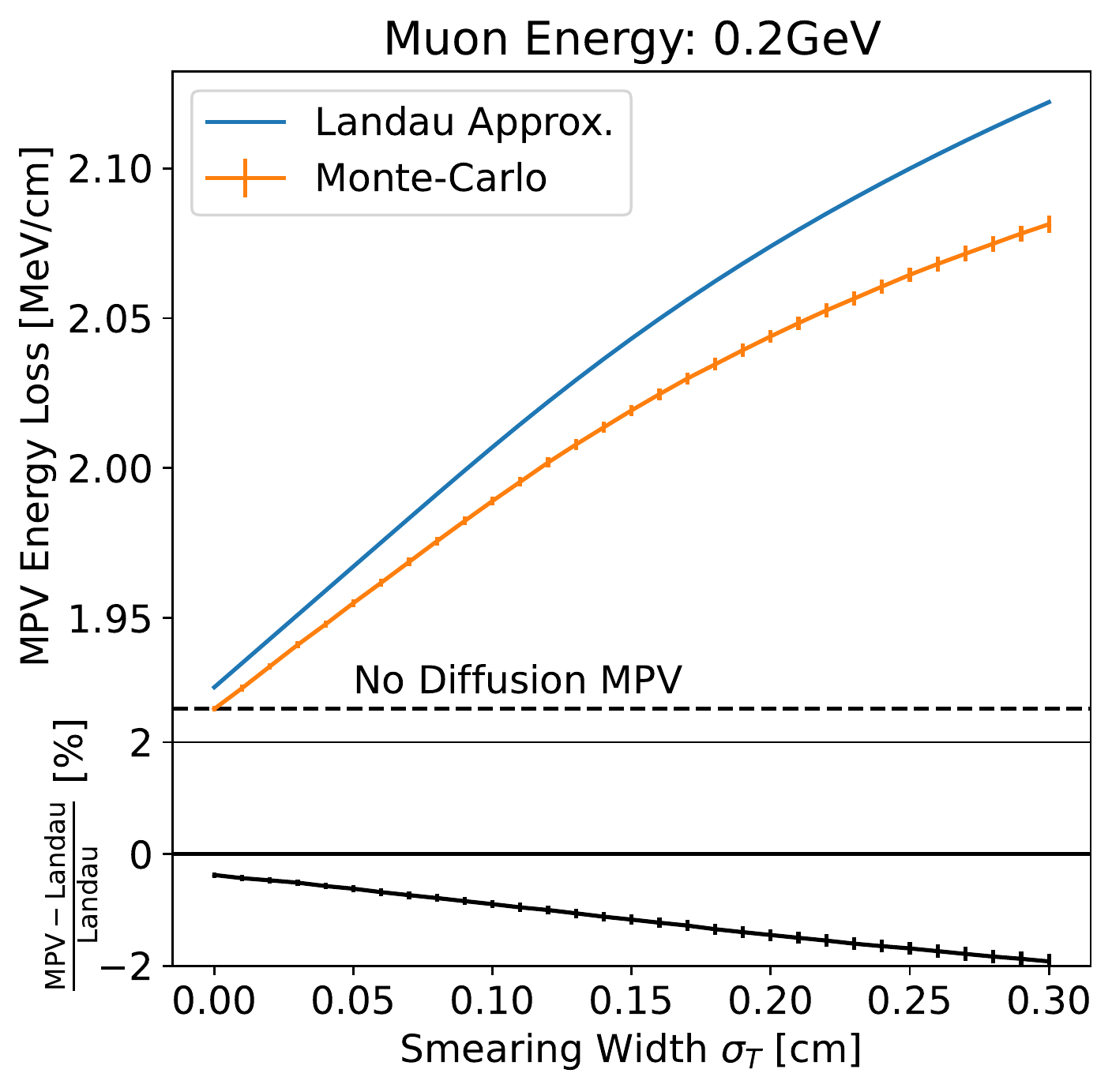}
	\includegraphics[width=0.325\textwidth]{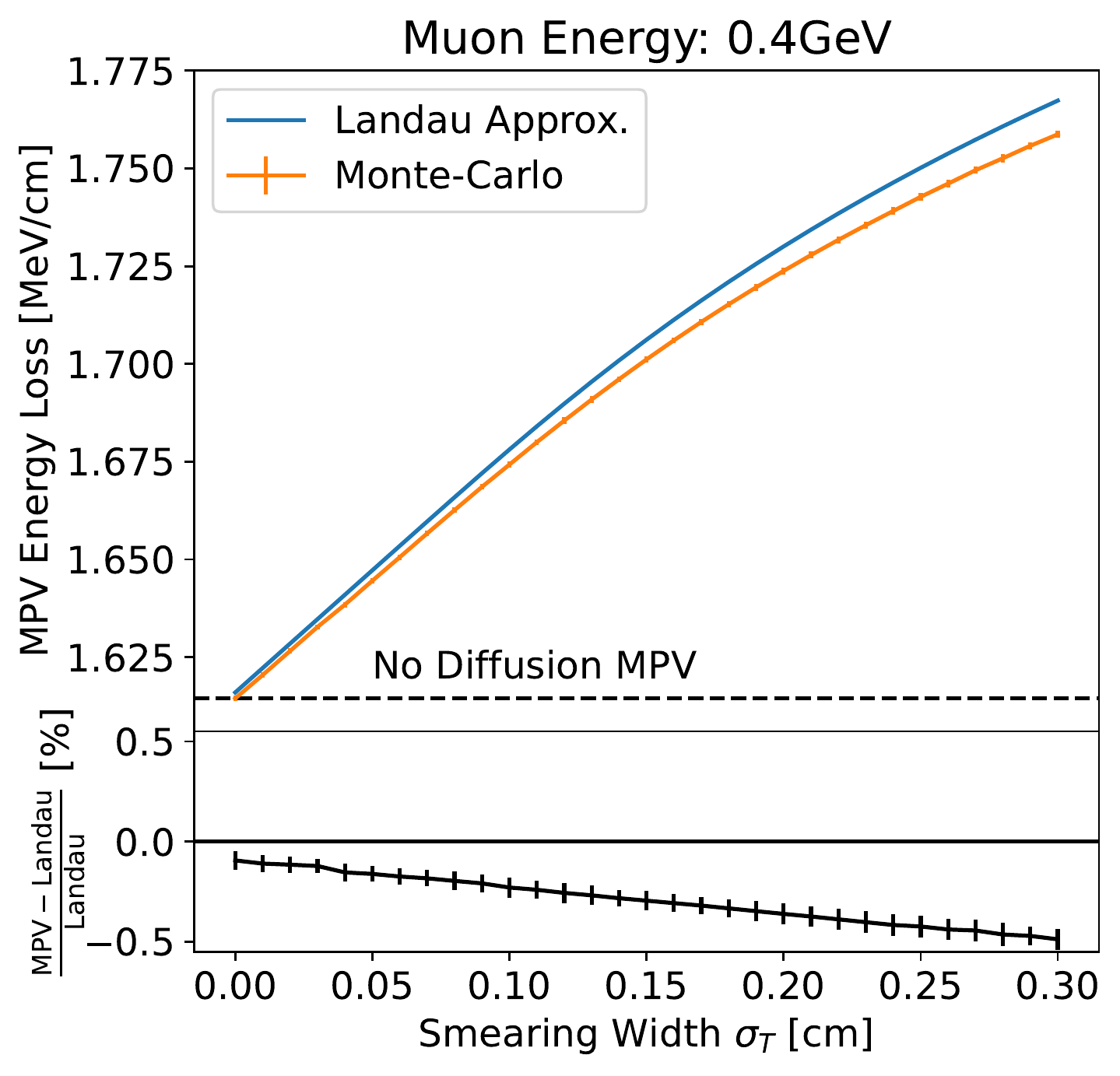}
	\includegraphics[width=0.325\textwidth]{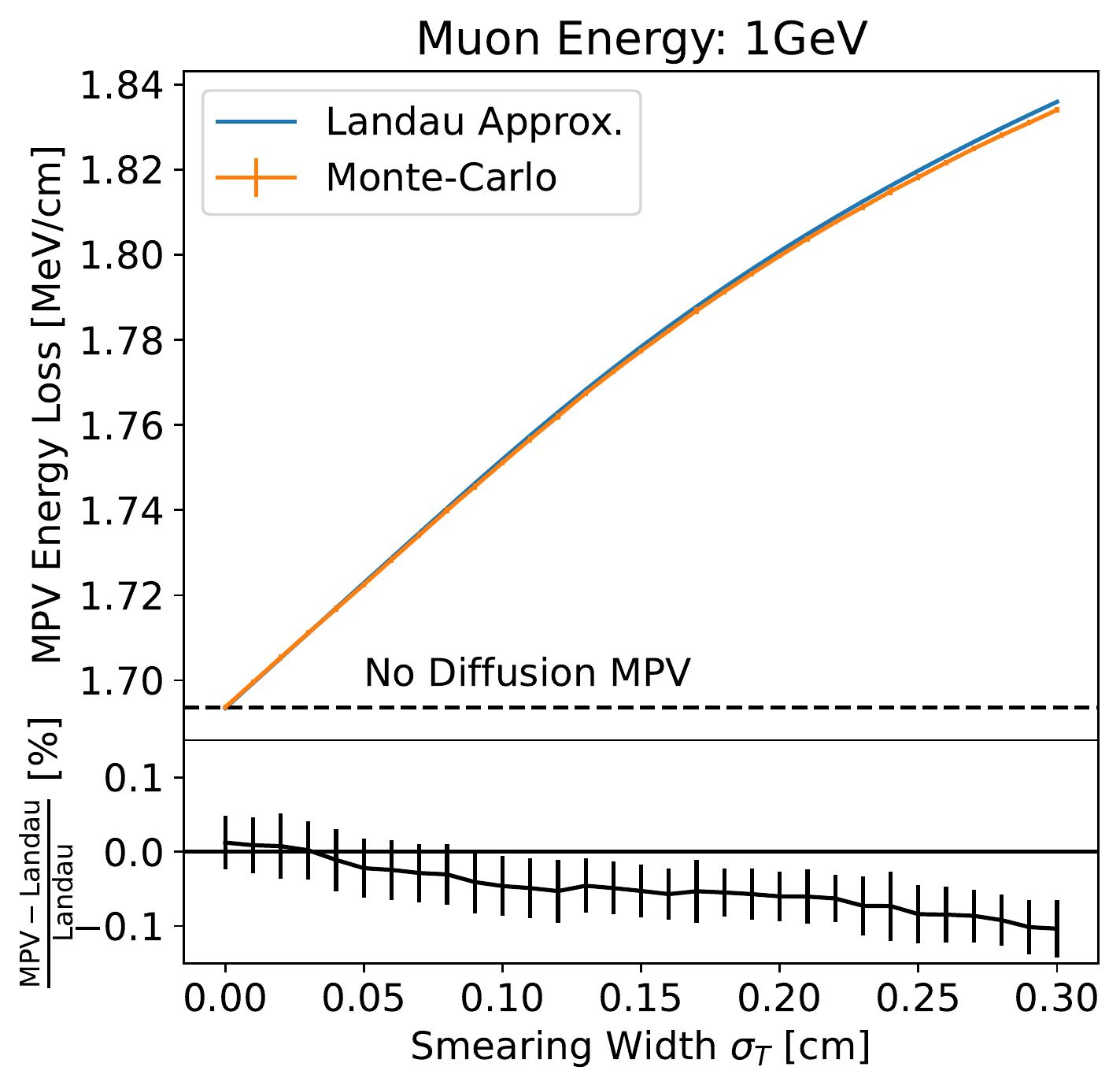}
	\caption{Comparison of the MPV energy loss from a Monte-Carlo simulation of the
	probability distribution in equation \ref{eq:LandauVavilov} (details in
	appendix \ref{sec:MC}) and the Landau MPV approximation (equation \ref{eq:landauMPV}) 
	for a channel ionization weight
	function given by a step
	function of width \SI{3}{mm} convolved with a Gaussian of varying widths (equation
	\ref{eq:wlartpc}). The comparison is
	made at various values of the muon energy. At low muon energy and large channel thickness,
	the thin-film/relativistic approximation underlying the Landau MPV prediction begins to
	breakdown, up to the percent level at \SI{0.2}{GeV}.}
	\label{fig:mpvcomp}
\end{figure}

\begin{figure}[t]
	\centering
	\includegraphics[width=0.75\textwidth]{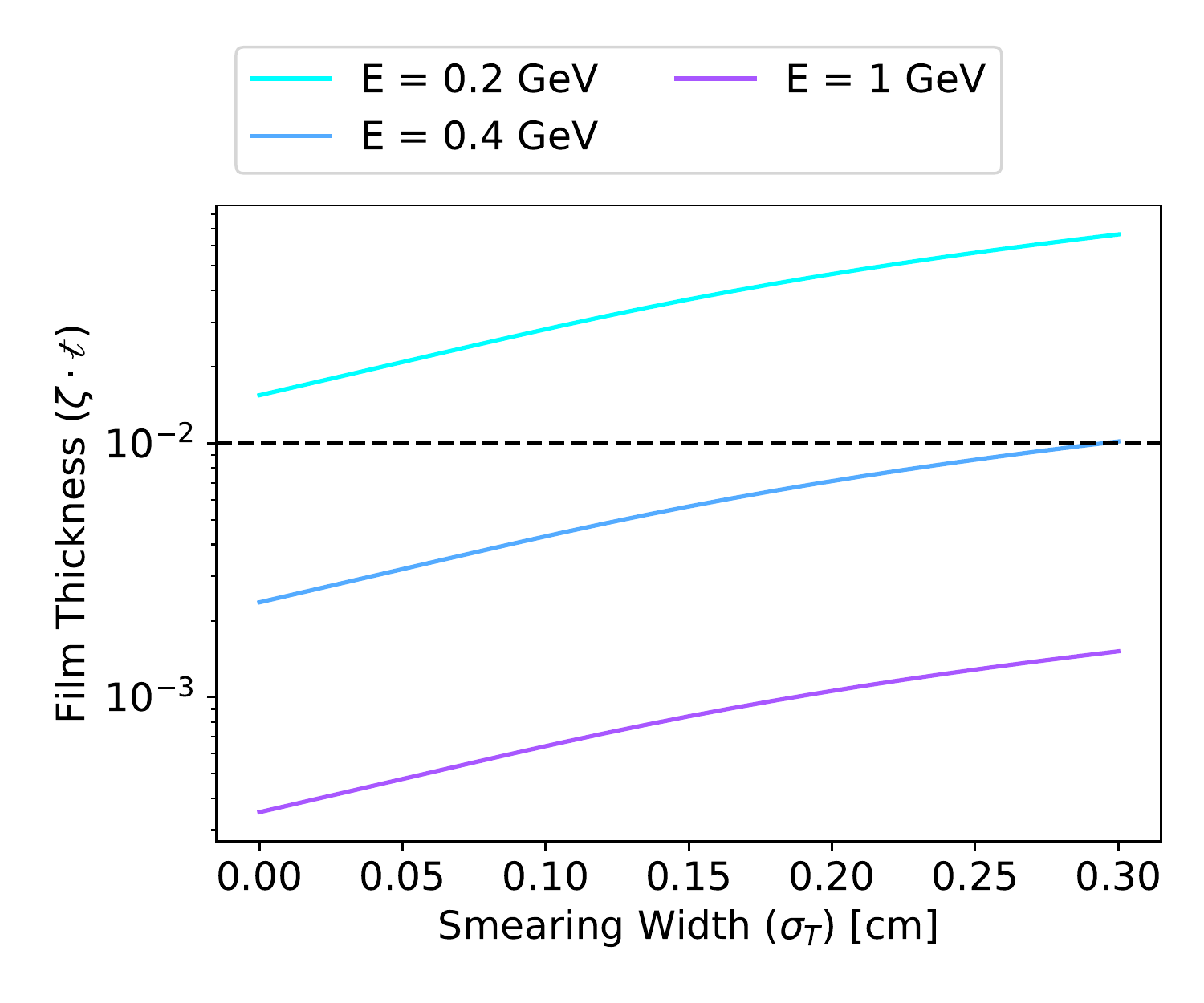} 
	\caption{Value of the ``film thickness'' ($\zeta\cdot\thicc$, $\zeta$ defined in
	equation \ref{eq:zeta} and $\thicc$ in equation \ref{eq:thicc}) for the muon energies
	and smearing values of figure \ref{fig:mpvcomp}.
	A dotted line is drawn at the crossover point above which the
	Landau distribution is generally considered to not be a valid approximation to the
	distribution of particle energy loss \cite{PDG}.}
	\label{fig:thickkappa}
\end{figure}

Figure \ref{fig:mpvcomp} plots the most-probable-value of energy loss obtained from a
Monte-Carlo simulation of muon energy loss observed by a LArTPC-like weight function, as a
function of the smearing width $\sigma_T$. 
(Implementation details are in appendix \ref{sec:MC}.)
The value is compared to the energy loss estimate 
from the thin-film approximation (i.e., using the value of the thickness in the Landau
distribution MPV
formula, equation \ref{eq:landauMPV}). The Landau approximation works well at small thickness and large muon energy. 
Outside of this region, numerically obtaining the peak of the general distribution with
the weight function (equation \ref{eq:LandauVavilov}) would be required.
Since the LArTPC weight function is not a step function, in this region
the distribution is also not a Landau-Vavilov distribution.
In the usual Vavilov case, one defines this region of phase space using the
parameter $\zeta \cdot \pitch$, which is a unit-less
measure of the ``film-thickness'' of a channel that is small for large particle energy and small
channel thicknesses.
Typically, the Landau distribution is taken
to be valid for $\zeta \cdot \pitch < 0.01$ \cite{PDG}. For our purposes, it is natural
to leverage $\zeta \cdot \thicc$. This value is plotted for the numerically computed
energies and thicknesses in figure \ref{fig:thickkappa}.

These same principles also apply to a LArTPC with a readout from an array of charge sensing
pixels. In this case, diffusion along both transverse directions is important. For a particle
moving in the ($\hat{y}, \hat{z}$) direction in the transverse plane that hits the pixel
at the location $(y_0, z_0)$, the weight function of
a square pixel with a side length $a$ as a
function of the particle position $r$ is
\begin{equation}
	\w_\text{pixel-LArTPC}(r) = \int\limits_0^a dy' \int\limits_0^a dz'
	\frac{1}{2\pi\sigma_\text{T}^2}
	e^{-(r\cdot \hat{y} - y_0 -y')^2/(2\sigma_\text{T}^2)}
	e^{-(r\cdot\hat{z} - z_0-z')^2/(2\sigma_\text{T}^2)} \,.
\end{equation}
The location that the track hits the pixel at is likely challenging to reconstruct. This is not
just a challenge for computing the weight function; the pitch of the track across the pixel
also depends on this location. One workaround would be to sum the reconstructed charge across
the pixels in either the $\hat{y}$ or $\hat{z}$ directions. In this case, the weight function
reduces to the wire plane readout case.

\section{Impact on LArTPC Calibration}
\label{sec:calib}

These findings impact LArTPC detector calibrations, which use particle energy loss
(especially from cosmic muons) as a ``standard candle''. 

\subsection{Energy Scale Calibration}

LArTPC detectors use muons to calibrate the map from digitized charge to energy. This is
done by calibrating the gain of the readout electronics (from ADC to number of electrons), applying any
detector non-uniformity factors, and then using external measurements of
the ionization work function \cite{Wion} and
recombination \cite{NEUTRecomb,ICARUSRecomb} to map from electrons to energy. The measurement of the gain is
done by measuring the distribution of ionization charge per pitch ($dQ/dx$) from muon energy
depositions binned in muon kinematics and
orientation to select for a single Landau distribution. A Landau function convolved with a
Gaussian function is
fit to this distribution, and the MPV of the fit is compared to the prediction from Bethe-Bloch
theory with the recombination model to extract the detector gain
(see prior calibrations: \cite{ProtoDUNEcalib,LArIATcalib,NEUTcalib,uBcalib}). 
This technique for energy scale calibration relies on the MPV energy loss of cosmic muons, which
depends on diffusion. Neglecting diffusion in the computation of this standard candle will bias
its use by an effect of a few percent.

As an example, in the MicroBooNE detector (where this effect is large due to the large drift time), for an
energy deposition at the cathode by a muon
with an energy of \SI{1}{GeV}, the energy loss ($dE/dx$) MPV is
\SI{1.79}{MeV\per cm}
accounting for diffusion in the channel thickness (using a \SI{0.3}{cm} wire pitch, 2.56m drift
length, \SI{0.1098}{cm\per\micro\second} drift velocity, and \SI{5.85}{cm\squared\per\second}
transverse diffusion constant reported in \cite{uBDL}).
The MPV would be \SI{1.69}{MeV\per cm} computed (incorrectly) using the track pitch, a
difference of 5.9\% (both of these
values neglect the density effect).
For further reference, table \ref{tbl:MPVs} lists the MPV $dE/dx$ for energy depositions from a
muon with \SI{1}{GeV} of energy
for a few LArTPC detector configurations.

In order to apply muon energy loss as a standard candle for energy scale calibration correctly, the effect of diffusion must be
considered. 
One approach would be to utilize only energy depositions near the anode at small drift time, 
where the effect of diffusion is small, though it could be challenging to acquire a sufficient 
statistical sample.
However, it is also possible to apply the
results of this work to incorporate all energy depositions in a consistent manner.
In the relativistic limit, this can be done by using equation \ref{eq:thicc} to
compute the thickness input to the Landau MPV $dE/dx$ equation. The thickness equation has four inputs:
the channel spacing, the transverse diffusion constant, the drift time, and the muon track
orientation. The channel spacing is a
constant of each detector. The transverse diffusion constant is a LAr property.
The drift time and track orientation can be reconstructed for each individual
energy deposition. Where the relativistic limit breaks down (i.e.\ where $\zeta\cdot\thicc >
0.01$), the general distribution of particle energy loss must be used (equation\
\ref{eq:LandauVavilov}). 
Since the LArTPC channel weight function is not a step function, in this region this
distribution is not a Landau-Vavilov distribution (see appendix \ref{subsec:vavilov} for a
proof). Thus, existing numerical routines such as the ROOT \cite{ROOT} \lstinline{VavilovAccurate}
function cannot be used.
For typical LArTPC detector configurations, the non-Landau region corresponds to the particle Bragg peak
(as shown by the values of $\zeta\cdot\thicc$ in figure \ref{fig:thickkappa}), so these
considerations are also important for modeling the distribution of particle energy losses for
particle identification.

\begin{table}[t]
\centering
\rowcolors{2}{gray!25}{white}
\hspace*{-0.5cm} \begin{tabular}{c | x{10mm} | x{10mm} | x{17mm} | x{23mm} | x{27mm} | x{14mm} }
	\rowcolor{gray!15}
	Detector & Wire Pitch [mm] & Drift Time [ms] & Diffusion Const. $D_T$ [\si{cm\squared\per s}] & MPV $dE/dx$, \quad No Diffusion
	[\si{MeV\per cm}] & MPV $dE/dx$ at \quad Cathode (Full Diff.) [\si{MeV\per cm}] &
	Difference [\%]\\
	\hline
	MicroBooNE \cite{MicroBooNEDet} & 3.00 & 2.33 & 5.85& 1.69 & 1.79 & 5.9\\
	ArgoNeuT \cite{ArgoNeuTD} & 4.00 &  0.295 & 12.0 (9.30) & 1.72 (1.72) & 1.76 (1.75) &
	2.3 (1.7)\\
	ICARUS \cite{SBNDet} & 3.00 & 0.960 & 12.0 (9.30)& 1.69 (1.69) & 1.78 (1.77) & 5.3
	(4.7)\\
	SBND \cite{SBNDet}& 3.00 & 1.28  & 12.0 (9.30) & 1.69 (1.69) & 1.79  (1.78) & 5.9 (5.3)\\
	DUNE-FD (SP) \cite{DUNEDet} & 4.71 &  2.2 & 12.0 (9.30) & 1.74 (1.74) & 1.82 (1.81) &
	4.6 (4.0)\\
\end{tabular}
\caption{
Values of the most-probable-value (MPV) of $dE/dx$ for a muon with \SI{1}{GeV} of energy with no 
perturbation from diffusion (this is equivalent to a deposition at the anode)
and at the cathode (where the effect of diffusion
is largest). The MicroBooNE transverse diffusion constant is taken from the extrapolated
measurement in that detector \cite{uBDL}, while the others are taken from
the Li et al.\ \cite{LiDL} (Atrazhev-Timoshkin \cite{Atrazhev}) prediction at an E-field of \SI{500}{V\per
cm}.
All MPVs neglect the density effect, which would
affect the diffusion on and off quantities equally.
}
\label{tbl:MPVs}
\end{table}

\subsection{Drift Direction Response Equalization}

The effect of diffusion can bias measurements attempting to equalize the response of the
detector as a function of drift time (removing effects such as LAr impurities). 
This is typically done by defining an observable of the $dQ/dx$
distribution from cosmic muons (such as the median \cite{uBcalib}) and applying a drift-dependent correction
factor to make the observable flat across the detector. However, since the underlying
distribution of energy loss is also drift-dependent in the presence of diffusion, these
procedures may be dividing-out a physical effect rather than just the detector response.
The effect of smearing induced by transverse diffusion is to push the value of the $dQ/dx$ MPV up
along the drift length, as opposed to LAr impurities, which push it down. 

\begin{figure}[t]
	\centering
	\includegraphics[width=0.75\textwidth]{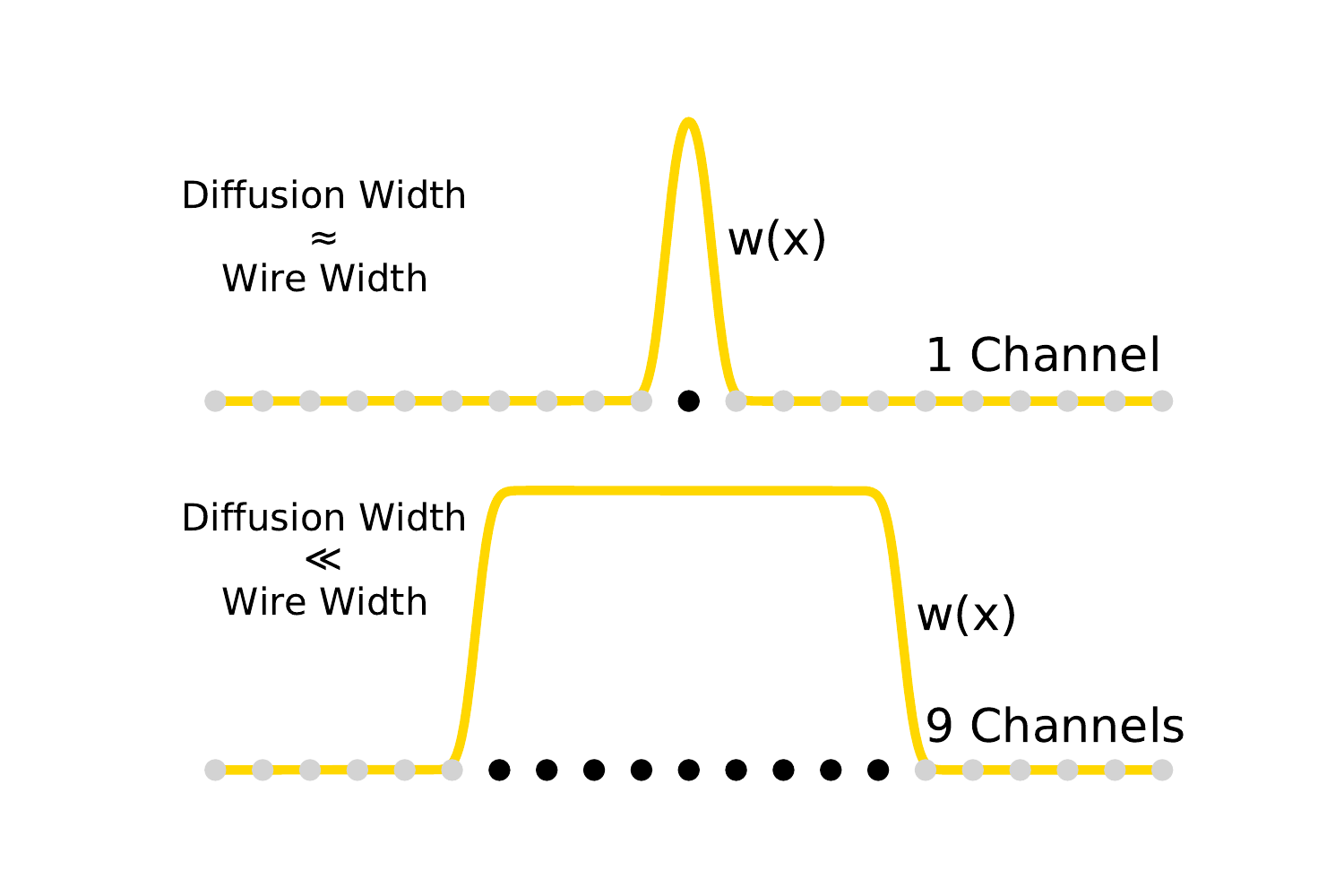}
	\caption{Measuring the charge wire-by-wire in a LArTPC (top) means that the thickness of
	the wire is significantly impacted by diffusion. Summing the charge across a large
	number of consecutive channels (bottom) diminishes this effect, which can be a useful
	technique for calibrations.}
	\label{fig:qsum}
\end{figure}

There are two strategies one can take to minimize the effect of diffusion on these measurements.
First, one can try to define a quantity of the $dQ/dx$ distribution that tracks to the mean
energy loss more than the MPV (such as a truncated mean). The mean value of energy loss is not
affected by diffusion. Second, one can ``coarse-grain'' the detector by summing hits across
many consecutive channels along the track into each value of $dQ/dx$. 
By combining enough channels, the length of the effective
channel spacing would dominate over the length of diffusion and thus the drift-dependent effect of
diffusion could be made negligible. This process is demonstrated diagrammatically in figure
\ref{fig:qsum}.

\subsection{Observation in Detectors}

Verifying the effect of diffusion on the peak value of particle energy loss in LArTPC detectors
may be challenging.
A number of different detector effects conspire to change the $dQ/dx$ by
drift time, all in their own way. One possibility could be to leverage the fact that this effect
only changes the MPV of particle charge depositions (as opposed to other effects which should change 
the mean and the MPV equally) and measure the ratio of a truncated mean to the MPV of $dQ/dx$ by
drift time.
Another would be to compare the ratio of a coarse-grained measurement of $dQ/dx$ (where
diffusion should be negligible) to a channel-by-channel measurement of $dQ/dx$. In any such
measurement, an experiment where other drift-dependent factors are small (small field
distortions and few LAr impurities) is ideal. 

The smearing which perturbs the channel thickness is induced by transverse diffusion, which has
never been measured directly in LAr at the electric fields typical in neutrino LArTPC detectors. 
Observation of this effect in LArTPC detectors in principle offers a method to measure transverse
diffusion. The change in the MPV energy loss depends on both muon kinematics and diffusion, so
such a measurement would also need to reconstruct the muon momentum.
Isolating the effect of diffusion on the MPV from other
detector effects will likely be quite challenging, and the impact of diffusion on the MPV is
indirect: large changes in the channel thickness only change the peak energy loss by a few percent. 
Still, a measurement of transverse diffusion (whether through this method or by looking directly
at how transverse diffusion impacts the width of ionization signals) is important for applying
the methods described in this paper to experiments.
Until it is measured directly, the value of transverse
diffusion represents a significant uncertainty for energy scale calibrations in a LArTPC. 

In terms of simulation, existing LArTPC experiments should be able to verify
in their Monte-Carlo simulations that this effect occurs. The change to the MPV is an emergent
effect of the Landau-Vavilov nature of particle energy loss combined with diffusion. Any LArTPC
detector simulation that includes these effects should output a $dE/dx$ distribution consistent
with the results of this work. (This effect has recently been demonstrated in a standalone LArTPC-like
simulation using \lstinline{GEANT4} in Ref. \cite{Michelle}.)

\section{Summary}

In this work, we have derived the distribution of particle energy loss recorded by a channel with a
position-dependent sensitivity to the particle ionization, described by a weight function $w(x)$.
In general, this distribution is
not equivalent to the Landau-Vavilov distribution, but in the thin-film limit it does converge
to a Landau distribution. In the Landau limit, 
the energy loss MPV can be computed using the thickness $\thicc = \pitch \hspace{0.05em}
e^{-\int \w(x)\text{log}[\w(x)]dx/\pitch}$, where $\pitch = \int \w(x)dx$. In a
LArTPC, $\w$
is given by the convolution of the step-function wire slice and the Gaussian effect of
transverse diffusion. In this case $\thicc \neq \pitch$. Prior calibrations
in LArTPCs have used the track pitch $\pitch$ as the input to the Landau MPV computation, which is
incorrect. In addition, since the effect of transverse diffusion is drift dependent, this effect
can bias attempts to equalize the detector response of a LArTPC along the drift direction.
Coarse-graining the detector is likely an effective method to mitigate this bias.
The value of transverse diffusion has never been quantitatively measured in LAr, and it
represents a significant systematic uncertainty on any energy scale calibration.

\acknowledgments

Thank you to Mike Mooney, Joseph Zennamo, and Ed Blucher for valuable conversations on the
content of this work. And thank you to Mike Mooney for a helpful discussion on the current state
of diffusion measurements. This material is based upon work supported by the National Science 
Foundation Graduate Research Fellowship under Grant No. DGE-1746045 and the 
National Science Foundation under Grant No. PHY-1913983.

\bibliographystyle{JHEP}
\bibliography{diffuse_mpv}

\appendix 
\section{The Distribution of Energy Loss Seen by a Channel with a Position-Dependent Sensitivity to
Particle Energy}
\label{sec:derivation}

\subsection{Derivation}
\label{subsec:derivation}

Typically, the Landau and Landau-Vavilov distributions are derived by means of a Laplace
transform leveraging the continuity property of the energy loss distribution
\cite{Vavilov:1957zz}. Here we take an alternative approach, applying the convolution property
(equation \ref{eq:conv}) by means of a Fourier transform. We also keep track of the channel
ionization weight function
as a function of the charged particle position $(\w(x))$.
We end at the same place, except for perturbations coming from the
weight function. To start, we discretize $\w(x)$ into weights $\w_i$ over infinitesimal steps
$dx_i$, and build up the probability distribution over the full weight function by performing a
product of convolutions:
\begin{equation}
	p_\w(E) = \int d T_1 \int d T_2 \cdots \int d T_n \frac{p_{dx_0}(\frac{E - T_n - \cdots -
	T_1}{\w_0})}{\w_0}
	\times
	\frac{p_{dx_1}(\frac{T_n - \cdots - T_1}{\w_1})}{\w_1} 
	\times \cdots \times
	\frac{p_{dx_n}(\frac{T_n}{\w_n})}{\w_n}
	\,.
\end{equation}
By use of a Fourier transform $\mathcal{F}$, we can turn this into a regular product:
\begin{linenomath}\begin{equation*}
	p_\w(E) = \underset{\tau\to E}{\mathcal{F}^{-1}} \ \underset{i}{\Pi} \
	\underset{T\to\tau}{\mathcal{F}} \frac{p_{dx_i}(T/\w_i)}{ \w_i}
	\,.
\end{equation*}\end{linenomath}
Applying the scale property of a Fourier Transform:
\begin{linenomath}\begin{equation*}
	p_\w(E) = \underset{\tau\to E}{\mathcal{F}^{-1}} \ \underset{i}{\Pi} \ \underset{T\to \w_i\tau}{\mathcal{F}} p_{dx_i}(T)
	\,.
\end{equation*}\end{linenomath}
By taking the exponential of the log of the RHS, we can manipulate it into a sum:
\begin{linenomath}\begin{equation*}
	p_\w(E) = \underset{\tau\to E}{\mathcal{F}^{-1}} \ \mathrm{exp} \left[ \sum\limits_i
	\mathrm{log} \underset{T\to \w_i\tau}{\mathcal{F}} p_{dx_i}(T)
	\right]
	\,.
\end{equation*}\end{linenomath}
Using the small$-\ell$ formula (\ref{eq:smalll}) for $p_{dx}(T)$, we can simplify its Fourier transform:
\begin{linenomath}\begin{equation*}
	p_\w(E) = \underset{\tau\to E}{\mathcal{F}^{-1}} \  \mathrm{exp} \left[ \sum\limits_i
	\mathrm{log} (1 - \sigma \rho dx_i +
	\rho dx_i \underset{T\to \w_i\tau}{\mathcal{F}} \frac{d\sigma}{dT})
	\right]
	\,.
\end{equation*}\end{linenomath}
By applying $\mathrm{log}(1 + \epsilon) \approx \epsilon$, we obtain:
\begin{linenomath}\begin{equation*}
        p_\w(E) = \underset{\tau\to E}{\mathcal{F}^{-1}} \ \mathrm{exp} \left[ \sum\limits_i -
	\sigma \rho dx_i +
        \rho dx_i \underset{T\to \w_i\tau}{\mathcal{F}} \frac{d\sigma}{dT}
        \right]
	\,.
\end{equation*}\end{linenomath}
Which can be neatly turned into an integral:
\begin{linenomath}\begin{equation*}
	p_\w(E) = \underset{\tau\to E}{\mathcal{F}^{-1}} \ \mathrm{exp} \left[ \int dx
	(-\rho\sigma + \rho \underset{T\to \w(x)\tau}{\mathcal{F}}
	\frac{d\sigma}{dT}) \right]
	\,.
\end{equation*}\end{linenomath}

Now we apply the definitions of $\sigma$ and $\mathcal{F}$ to evaluate the integral. Noting
$d\sigma/dT \neq 0$ only for $0 < T < T_\text{max}$, we obtain
\begin{linenomath}\begin{equation*}
	\begin{split}
	p_\w(E) &= \underset{\tau\to E}{\mathcal{F}^{-1}} \ \mathrm{exp} \left[
	\rho \int dx \int\limits_{0}^{T_\text{max}} dT \frac{d\sigma}{dT} \left(e^{-2\pi i \tau
	\w(x) T} - 1\right)\right] \\
	&= \underset{\tau\to E}{\mathcal{F}^{-1}} \ \mathrm{exp} \left[
	\rho \frac{2\pi r_e^2 m_e}{\beta^2} \int dx \int\limits_{0}^{T_\text{max}} dT
	\frac{1-\beta^2 T/T_\text{max}}{T^2} \left(e^{-2\pi i \tau
        \w(x) T} - 1\right)\right]
	\,,
	\end{split}
\end{equation*}\end{linenomath}
where we have used the formula for the bare cross section (equation \ref{eq:xsec}) in place of
$d\sigma/dT$. 

The integrand diverges as $1/T$ as $T\to 0$. This divergence appears because we have used the
cross section of scattering on bare electrons instead of atomic electrons. At low energy
transfer the cross section is modified by atomic effects, which in particular
cutoff the cross section near the excitation energy (above $T=0$) to remove the divergence.
We can remove the divergent behavior of the integrand by adding and subtracting the mean energy loss
$\overline{E} = \int dx \int dT \rho \frac{d\sigma}{dT} T \w(x)$.
That subtracting the mean energy loss makes the integrand converge indicates that the shape of
the distribution is not sensitive to atomic effects; these only act to change the
mean. So once the impact of atomic effects on the mean energy loss is
accounted for, it is safe to apply the bare cross section to find the shape of the distribution.
Thus, for the purpose of this derivation, we take the mean energy
loss as an external input (from Bethe-Bloch theory \cite{Bethe}) and find how the mean relates to the shape
of the distribution. (There is also nothing new here coming from the weight function; this is
precisely the same approximation that is made by Vavilov \cite{Vavilov:1957zz}). 
Applying this substitution, we get
\begin{equation}
	p_\w(E) = \underset{\tau\to E}{\mathcal{F}^{-1}} \ \mathrm{exp} \left[
	-2\pi i \overline{E}\tau +\rho \frac{2\pi r_e^2 m_e}{\beta^2} \int dx \int\limits_{0}^{T_\text{max}} dT
        \frac{1-\beta^2 T/T_\text{max}}{T^2} \left(e^{-2\pi i \tau
        \w(x) T} + 2\pi i \tau \w(x) T - 1\right)\right]
	\label{eq:GeneralLV}
	\,.
\end{equation}
This integral converges to
\begin{linenomath}\begin{equation*}
	\begin{split}
	\hspace{-2em} p_\w(E) = \underset{\tau\to E}{\mathcal{F}^{-1}} \ \mathrm{exp} \left[
-2\pi i \overline{E}\tau + \rho \vphantom{\int} \right. &\frac{2\pi r_e^2 m_e}{\beta^2
T_\text{max}} \int dx
\left( \vphantom{e^{blah}}\right. 1 - e^{-2 i \pi T_\text{max} \w(x) \tau} - 2\pi i \tau
\w(x) T_\text{max}(1 + \beta^2) +
\\
& (\beta^2 + 2
i \pi T_\text{max} \w(x) \tau)(-\text{Ei}[-2 i \pi T_\text{max} \w(x) \tau] + \text{log}[2 i
\pi T_\text{max} \w(x) \tau] + \gamma_\text{EM})
\left. \vphantom{e^{blah}}\right)
\left. \vphantom{\int}\right]
\,,
\end{split}
\end{equation*}\end{linenomath}
where $\gamma_\text{EM}$ is the Euler constant and $\text{Ei}$ is the exponential integral function,
$\text{Ei}(x) = -\int\limits_{-x}^\infty dt \ e^{-t}/t$. 
Next, we expand the inverse Fourier transform $\mathcal{F}^{-1}$:
\begin{linenomath}\begin{equation*}
\begin{split}
	\hspace{-2em} p_\w(E) = \int\limits_{-\infty}^{\infty} 
d\tau \ \mathrm{exp} \left[2\pi i E \tau
\vphantom{\int} \right. &
-2\pi i \overline{E}\tau + \rho\frac{2\pi r_e^2 m_e}{\beta^2 T_\text{max}} \int dx
\left( \vphantom{e^{blah}}\right. 1 - e^{-2 i \pi T_\text{max} \w(x) \tau} - 2\pi i \tau
\w(x) T_\text{max}(1 + \beta^2) +
\\
& (\beta^2 + 2
i \pi T_\text{max} \w(x) \tau)(-\text{Ei}[-2 i \pi T_\text{max} \w(x) \tau] + \text{log}[2 i
\pi T_\text{max} \w(x) \tau] + \gamma_\text{EM})
\left. \vphantom{e^{blah}}\right)
\left. \vphantom{\int}\right]
\,.
\end{split}
\end{equation*}\end{linenomath}
To simplify these integrals, we can leverage the $\zeta$ quantify defined earlier (equation
\ref{eq:zeta})
and
$z \equiv 2\pi \tau \zeta T_\text{max}$:
\begin{equation}
\begin{split}
	p_\w(E) = \frac{1}{2\pi\zeta T_\text{max}} \int\limits_{-\infty}^{\infty} dz \ \mathrm{exp}
\left[\frac{iz}{\zeta T_\text{max}}
\vphantom{\int} \right. &
(E-\overline{E}) 
+ \int dx  \ \zeta ( 1
 - e^{-i \w(x) z/\zeta}) 
 - i z \w(x) (1 + \beta^2) + \\
& (\zeta\beta^2 + i \w(x) z)(-\text{Ei}[-i \w(x) z/\zeta] +
\text{log}[ i \w(x) z/\zeta] + \gamma_\text{EM})
\left. \vphantom{\int} \right] 
\label{eq:LandauVavilovApp}
\,.
\end{split}
\end{equation}

This integral definition gives the general result of the probability distribution of energy loss
observed by some channel with a position-dependent weight function $\w(x)$. In the nominal case, we
would
replace $\int dx \to \pitch$ for some channel pitch $\pitch$ and would obtain the Landau-Vavilov distribution. 
From here, we will consider for which channel sensitivities the distribution
is equal to the Landau distribution (in the thin film case) or the Landau-Vavilov distribution
(in the general case). In section \ref{subsec:landauapp} it is shown that for all channel sensitivities that
satisfy the thin film approximation, the distribution is a Landau one. Finally, in section
\ref{subsec:vavilov} we find that only for specific channel sensitivities is the distribution
equivalent to a Landau-Vavilov distribution in the general case.

\subsection{The Landau Limit}
\label{subsec:landauapp}

To restrict to the Landau case, we take the thin-film approximation.
In the usual derivation, one takes the
limit that $\zeta \cdot \ell \ll 1$, where $\ell$ is the width of the step function.
In our case, since we don't have a single such width, we have to be more careful
about this approximation. In this case we can make a requirement on $\w$ -- that the range of
values where $\w$ is not $\ll 1$, $r$, has the property that $\zeta\cdot r \ll 1$. 
Then, inside the
integrand of $\int dx$, we can take the limit that $\zeta$ is small. In this limit, 
$p_\w(E)$ converges to
\begin{equation}
        p_\w(E) = \frac{1}{2\pi i \zeta T_\text{max} \pitch}
        \int\limits_{-i\infty}^{i\infty} dz' \ \text{exp} \left[
        z' \left(
	\lambda
        + \text{log}|z'|\right)\right]
        \,,
\end{equation}
where $z' = i z / \zeta T_\text{max}$, $\pitch\equiv \int \w(x)dx$, and $\lambda = \frac{E - \overline{E}}{\zeta
T_\text{max}\pitch} - \text{log}\zeta \pitch
+ \gamma_\text{EM} - 1 - \beta^2 + \frac{\int dx \ \w(x) \text{log}[\w(x)]}{\pitch}$.
This equation can be recognized as the Landau distribution for a parameter $\lambda$.

\subsection{General / Landau-Vavilov Case}
\label{subsec:vavilov}

To understand the general case, we will examine the cumulants of the probability distribution. To do this, it is useful to go back
to the definition in equation \ref{eq:GeneralLV}, modified slightly so that we obtain the
cumulant-generating function $K(\tau) = \text{log}\ \text{E}\left[e^{-i \tau E}\right]$:
\begin{equation}
        K(\tau) = 
        -i \overline{E}\tau +\zeta T_\text{max} \int dx \int\limits_{0}^{T_\text{max}} dT
        \frac{1-\beta^2 T/T_\text{max}}{T^2} \left(e^{-i \tau
        \w(x) T} + i \tau 
	\w(x) T - 1\right)
	\,.
\end{equation}
We expand the term in parentheses in a Taylor series:
\begin{linenomath}\begin{equation*}
        K(\tau) =
        -i \overline{E}\tau +\zeta T_\text{max} \int\limits_{0}^{T_\text{max}} dT
	\frac{1-\beta^2 T/T_\text{max}}{T^2} \sum\limits_{n=2}^\infty
	\frac{(-i \tau T)^n}{n!} \int dx \ \w(x)^n
	\,.
\end{equation*}\end{linenomath}
Which can be simplified to
\begin{linenomath}\begin{equation*}
        K(\tau) =
	-i \overline{E}\tau +\zeta T_\text{max} \sum\limits_{n=2}^\infty
	\frac{(-i \tau)^n}{n!}
	T_\text{max}^{n-1}\left(\frac{1}{n-1} - \frac{\beta^2}{n}\right)
	\int dx \ \w(x)^n
	\,.
\end{equation*}\end{linenomath}
From here, the nth cumulant $\kappa_n = i^nK^{(n)}(0)$ can be
directly read off:
\begin{equation}
	\begin{split}
		\kappa_1 &= \overline{E}\\
		\kappa_n &= \zeta T_\text{max}^n\left(\frac{1}{n-1} - \frac{\beta^2}{n}\right)
        \int dx \ \w(x)^n
	\label{eq:cumulant}
	\,.
	\end{split}
\end{equation}

The cumulants of the Landau-Vavilov distribution are given for $\w(x) = \Theta(x) -
\Theta(x+\pitch)$
for some pitch $\pitch$. Thus, the cumulants of the Landau-Vavilov distribution are
\begin{equation}
	\begin{split}
		\kappa^\text{LV}_1 &= \overline{E}\\
		\kappa^\text{LV}_n &= \zeta \pitch T_\text{max}^n\left(\frac{1}{n-1} - \frac{\beta^2}{n}\right)
	\label{eq:LVcumulant}
	\,.
	\end{split}
\end{equation}
A necessary but not sufficient condition for two probability distributions to be equivalent is
that they have the same cumulants. Allowing for the distributions to be different by location
and scale parameters, the n-th cumulant must be equal up to a multiplicative (scale) factor
$c^n$. Thus, we need $\kappa_n = \kappa^\text{LV}_n c^n$ for a distribution (with cumulants
$\kappa$) to be equivalent to the Landau-Vavilov distribution up to location and scale
parameters. This puts a requirement on $w$ that $\int dx \ \w(x)^n = \pitch c^{n-1}$ for all
integers $n \geq 1$ for some constant $c$ and the pitch $\pitch$. 

The cumulants being equivalent is not by itself a sufficient condition for the probability
distributions to be the same. However, given this property on $\w$ we can simplify the
distribution further -- starting from equation \ref{eq:GeneralLV}:
\begin{linenomath}\begin{equation*}
	\begin{split}
	p_\w(E) &= \underset{\tau\to E}{\mathcal{F}^{-1}} \ \mathrm{exp} \left[
	-2\pi i \overline{E}\tau +\rho \frac{2\pi r_e^2 m_e}{\beta^2} \int dx \int\limits_{0}^{T_\text{max}} dT
        \frac{1-\beta^2 T/T_\text{max}}{T^2} \left(e^{-2\pi i \tau
        \w(x) T} + 2\pi i \tau \w(x) T - 1\right)\right]\\
	&= \underset{\tau\to E}{\mathcal{F}^{-1}} \ \mathrm{exp} \left[
	-2\pi i \overline{E}\tau +\rho \frac{2\pi r_e^2 m_e}{\beta^2} \frac{\pitch}{c} \int\limits_{0}^{T_\text{max}} dT
        \frac{1-\beta^2 T/T_\text{max}}{T^2} \left(e^{-2 \pi c i \tau
        T} + 2\pi c i \tau T - 1\right)\right]
	\,,
	\end{split}
\end{equation*}\end{linenomath}
which integrates to:
\begin{linenomath}\begin{equation*}
	\begin{split}
	\hspace{-2em} p_\w(E) = \underset{\tau\to E}{\mathcal{F}^{-1}} \ \mathrm{exp} \left[
-2\pi i \overline{E}\tau + \rho \vphantom{\int} \right. &\frac{2\pi r_e^2 m_e}{\beta^2
T_\text{max}} \frac{\pitch}{c}
\left( \vphantom{e^{blah}}\right. 1 - e^{-2 i \pi T_\text{max} c \tau} - 2\pi i \tau
c T_\text{max}(1 + \beta^2) +
\\
& (\beta^2 + 2
i \pi T_\text{max} c \tau)(-\text{Ei}[-2 i \pi T_\text{max} c \tau] + \text{log}[2 i
\pi T_\text{max} c \tau] + \gamma_\text{EM})
\left. \vphantom{e^{blah}}\right)
\left. \vphantom{\int}\right]
\,.
\end{split}
\end{equation*}\end{linenomath}
Then, defining $\zeta' = \rho \frac{2\pi r_e^2m_e}{T_\text{max} \beta^2}\frac{\pitch}{c} $
and
$z' = 2\pi i \tau T_\text{max} c$:
\begin{linenomath}\begin{equation*}
\begin{split}
p_\w(E) = \frac{1}{2\pi i T_\text{max} c} \int\limits_{-i\infty}^{i\infty} dz' \ \mathrm{exp}
\left[\frac{z'}{T_\text{max} c}
\vphantom{\int} \right. &
(E-\overline{E}) 
+ \zeta' ( 1
 - e^{-z'}) 
 - z' \zeta' (1 + \beta^2) +\\
& \zeta'(\beta^2 +  z')(-\text{Ei}[-z'] +
\text{log}[z'] + \gamma_\text{EM})
\left. \vphantom{\int} \right] 
\,,
\end{split}
\end{equation*}\end{linenomath}
which is the Landau-Vavilov distribution with a scale parameter $c$ (this can be verified
against \cite{Vavilov:1957zz} equation 8, with somewhat different notation). Thus, the probability
distribution of energy loss recorded by a channel with a weight function $\w(x)$ is equal to the Landau-Vavilov
distribution precisely when
\begin{equation}
	\int dx \ \w(x)^n = \pitch c^{n-1}
	\label{eq:wprop}
\end{equation}
for all integers $n \geq 1$ for some $\pitch, c$.

We can show further that this requirement means $\w(x)$ is equal to some number of
non-overlapping step functions
multiplied by a scale factor $c$. First, define $\pitch' = c\pitch$ and $\w'(x) =
\w(x)/c$. Then \ref{eq:wprop} being satisfied means $\int dx \  \w'(x)^n = \pitch'$ for all
$n\geq 1$. We show this means $\w'(x)$ is equal to 1 or 0 for all x. Take $n$ large enough that
for $\w' < 1$, $\w'^n \approx 0$ and for $\w' > 1$, $\w'^n \approx \infty$. Then $\w'(x)$ can't have some
compact region where $\w' > 1$, or else the integral would diverge. In this case, the
integral breaks down to a sum of the compact regions where $\w' = 1$: $\underset{n\to
\infty}{\mathrm{lim}} \int dx \ \w'(x)^n = \sum\limits_\mathrm{region-i} r_i = \pitch'$, where $r_i$ is the
length of region $i$ where $\w' = 1$. Then, consider the integral $\int dx \ w'(x)$. This is equal to those same
regions plus the integral of $\w'(x)$ outside those regions: $\int dx \ \w'(x) =
\sum\limits_\mathrm{region-i} r_i + \int\limits_{\w'\neq 1} \w'(x)$. Since
$\sum\limits_\mathrm{region-i} r_i = \pitch'$, we need $\int\limits_{\w'\neq 1} \w'(x) = 0$.
Since $\w'$ is positive for all $x$, this means $\w'$ must be equal to 0 wherever $\w'\neq 1$.
This means that $\w'$ should be given by the sum of some number of non-overlapping step
functions at location $x_i$ of length $a_i$: $\w'(x) = \sum_{i} \Theta(x - x_i) - \Theta(x - x_i
- a_i)$. Translating this back to $\w$, this means $\w$ must be given by the sum of those
step functions multiplied by some constant $c$ where $0 \leq c \leq 1$.

Thus, when not considering the relativistic limit, the distribution of particle energy loss
recorded
by a channel with a weight function $\w(x)$ is only equal to a Landau-Vavilov distribution (up to location
and scale factors) when $\w(x)$ is given by the sum of step functions multiplied by some overall
constant. In general, the probability distribution will be different and is given by equation
\ref{eq:LandauVavilovApp} (equation \ref{eq:LandauVavilov} in the main text). 
The cumulants of this distribution are given by equation \ref{eq:cumulant}.

\section{Implementation Details of the Monte Carlo}
\label{sec:MC}
The theoretical results of this note have been augmented by a simple Monte-Carlo simulation of
muon energy loss recorded by a LArTPC-like channel weight function. 
The values of the parameters of the computation (explained below) are provided
in table \ref{tbl:MCval}.
The Monte Carlo simulation consisted of sampling
energy loss for muons in steps of $\Delta x$ over a total length $\ell_\text{MC}$, 
then summing up the energy loss from a weight function $\w(x)$ computed at each $\Delta x$ step. 
The step-size $\Delta x$ was chosen so that the weight function did not appreciably vary across
it.
At each $\Delta x$ step, the distribution of energy loss was modeled by a Landau-Vavilov
distribution with the parameters of the muon kinematics and LAr properties as input. 


At the $\Delta x$ value required by this simulation, the Vavilov-$\kappa$ parameter of 
the Landau-Vavilov distribution was too small to be computed by widely available libraries. Thus
we implemented our own computation of the energy loss distribution. First, we computed the
small-length distribution of energy loss (using equation \ref{eq:smalll}) over a distance $\epsilon$.
This distribution was computed on a grid of $N_\text{MC}$ points up to an energy cutoff
$E_\text{max}$.
To incorporate the effect of atomic effects on the cross section, we cut off the bare cross
section at the mean ionization energy $I_0$ and added a term of constant energy loss to fix the value of mean
energy loss. This is a valid procedure because, as discussed in appendix
\ref{sec:derivation}, the shape of the Landau-Vavilov distribution is independent of atomic
effects as long as the mean is correct. After computing the small-length energy loss
distribution, we performed a series of discrete convolutions, doubling the length of the
probability distribution at each iteration, until we obtained the desired $\Delta x$. 

\begin{figure}[t]
	\centering
	\includegraphics[width=0.75\textwidth]{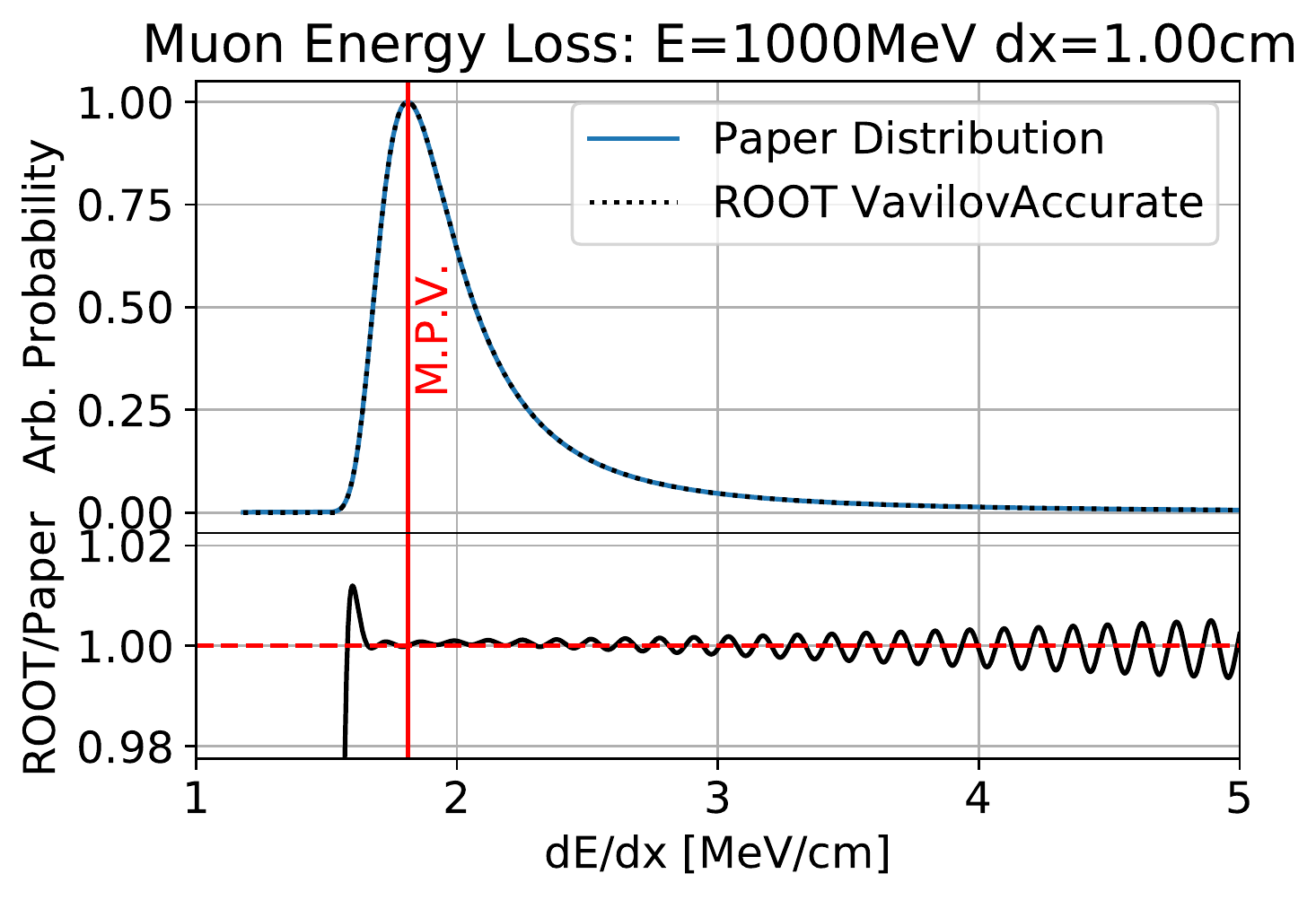}
	\caption{Validation of the Landau-Vavilov distribution implemented for this paper
	compared to the ROOT \lstinline{VavilovAccurate} implementation for a muon with
	\SI{1}{GeV} of
	energy for a thickness of \SI{1}{cm}.}
	\label{fig:validateVavilov}
\end{figure}

To validate this procedure,
we also compared the energy loss distribution to the ROOT 
\cite{ROOT} Landau-Vavilov function \lstinline{ROOT::Math::VavilovAccurate} at a length it was
able to compute for the simulated energies (\SI{1}{cm}): see figure \ref{fig:validateVavilov}.
Differences between the paper and the ROOT Landau-Vavilov distributions are on the order of a
tenth of a percent, and are smaller in the region of the distribution near the peak.

\begin{figure}[t]
	\centering
	Muon Energy: \SI{0.2}{GeV}
	\hspace{3cm}
	Muon Energy: \SI{0.4}{GeV}\\
	\includegraphics[width=0.22\textwidth]{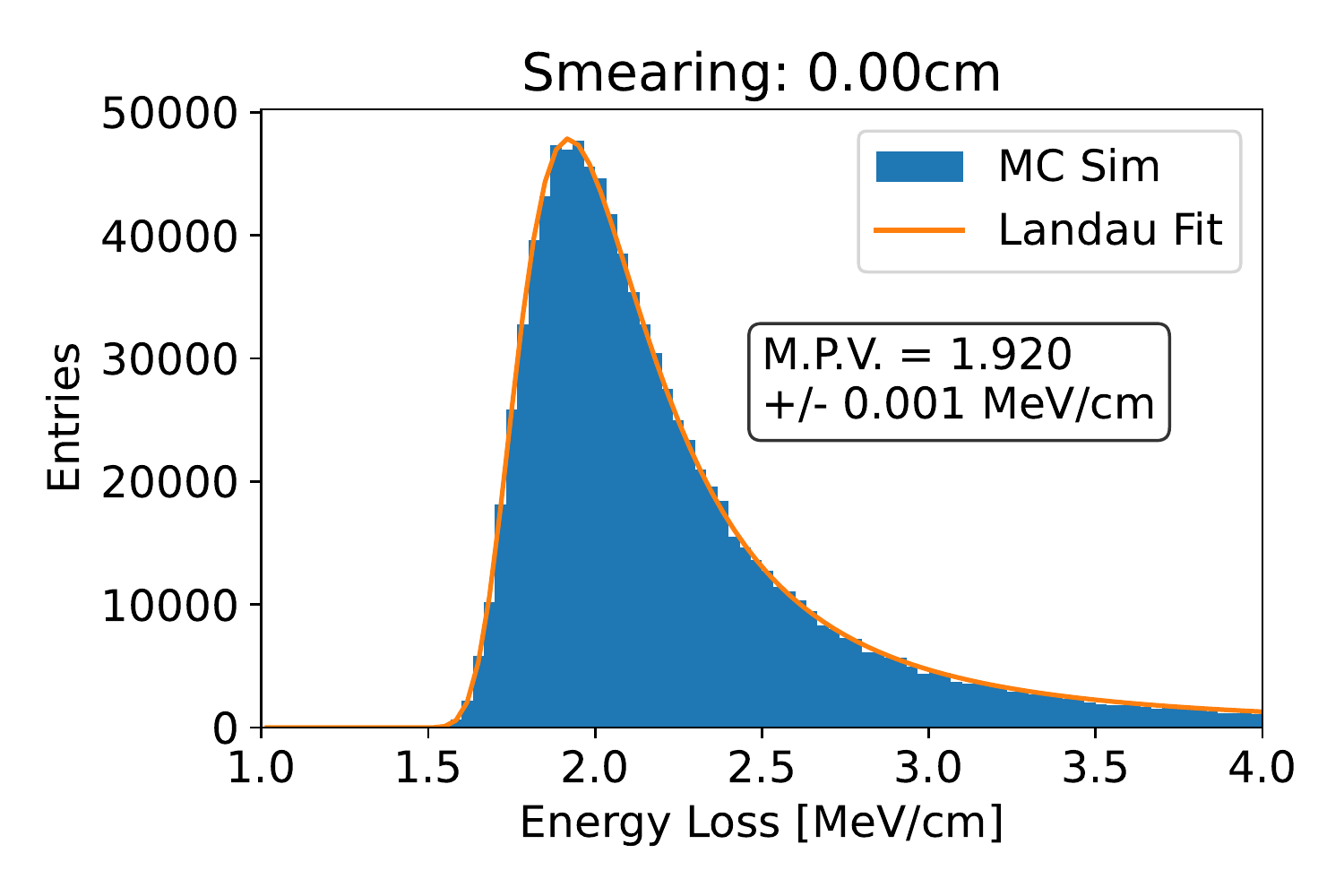}
	\includegraphics[width=0.22\textwidth]{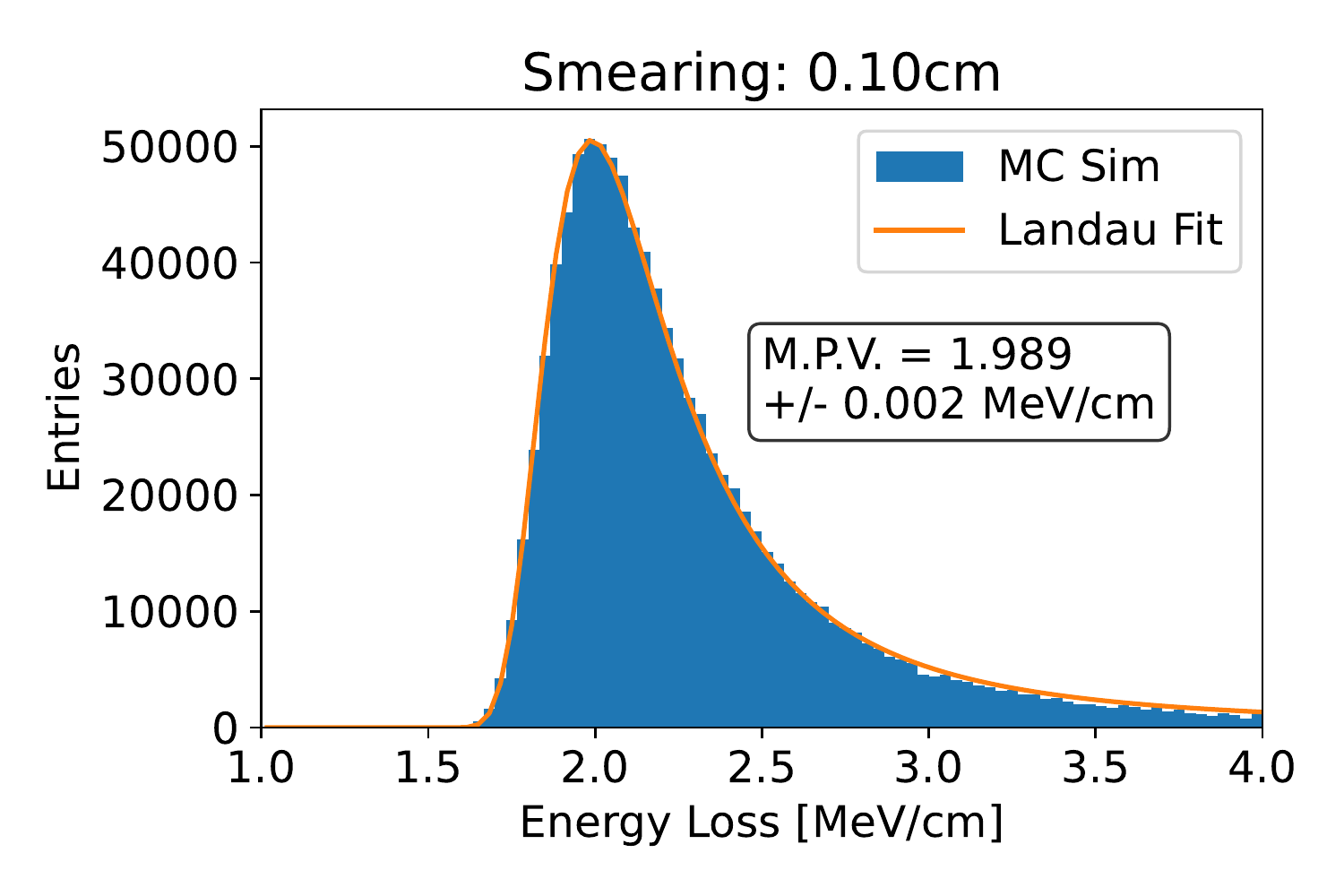}
	\hspace{1cm}
	\includegraphics[width=0.22\textwidth]{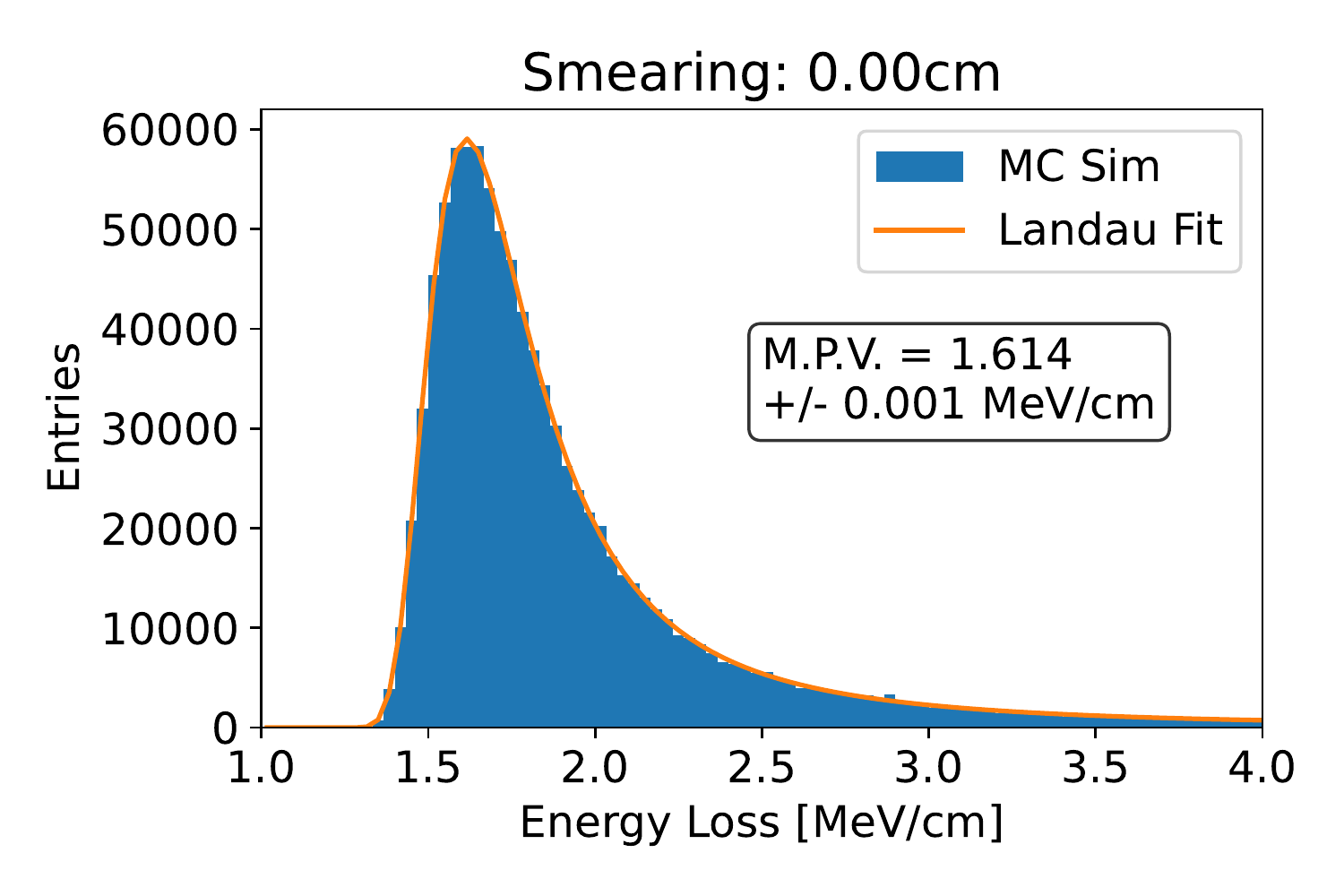}
	\includegraphics[width=0.22\textwidth]{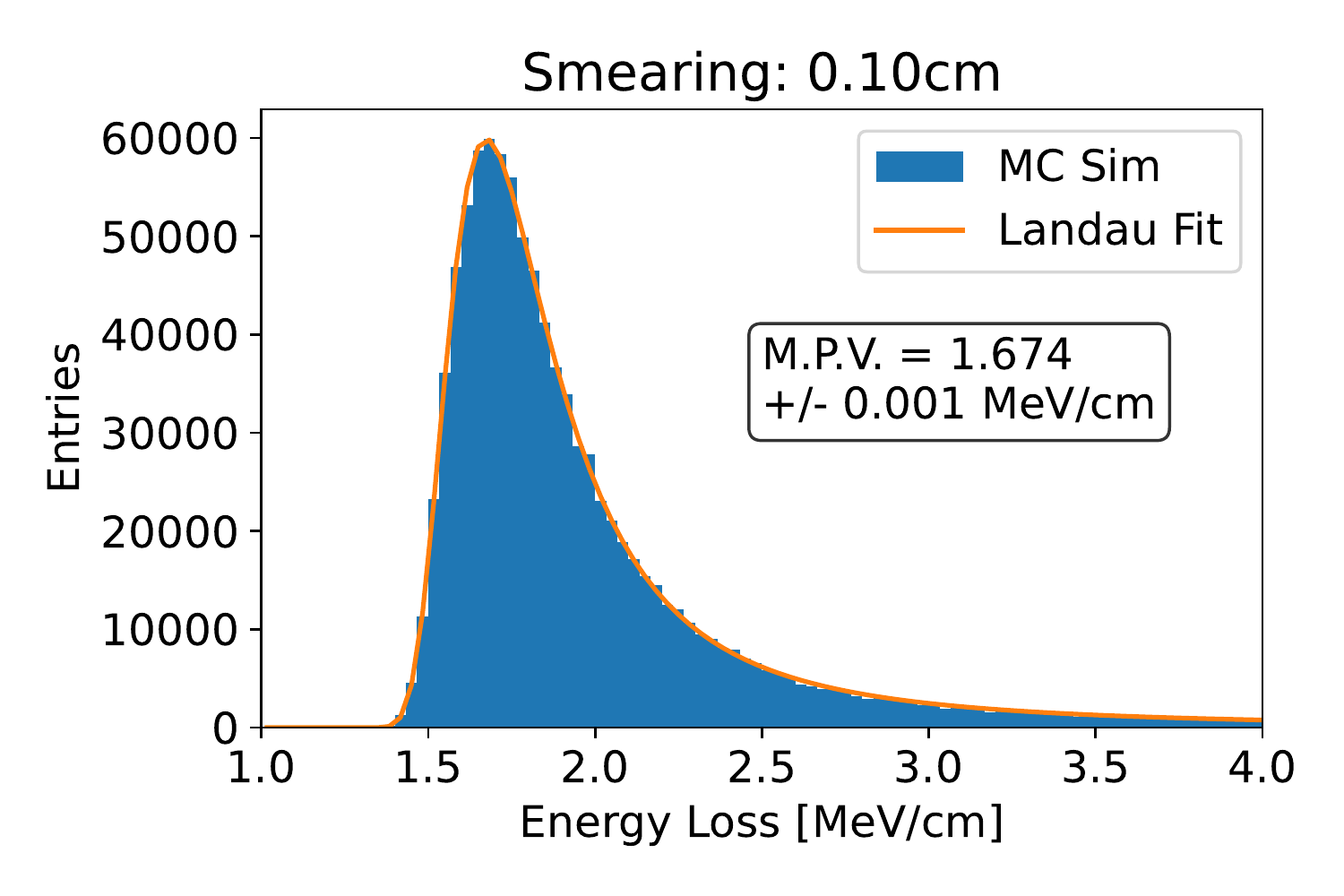}
	
	\includegraphics[width=0.22\textwidth]{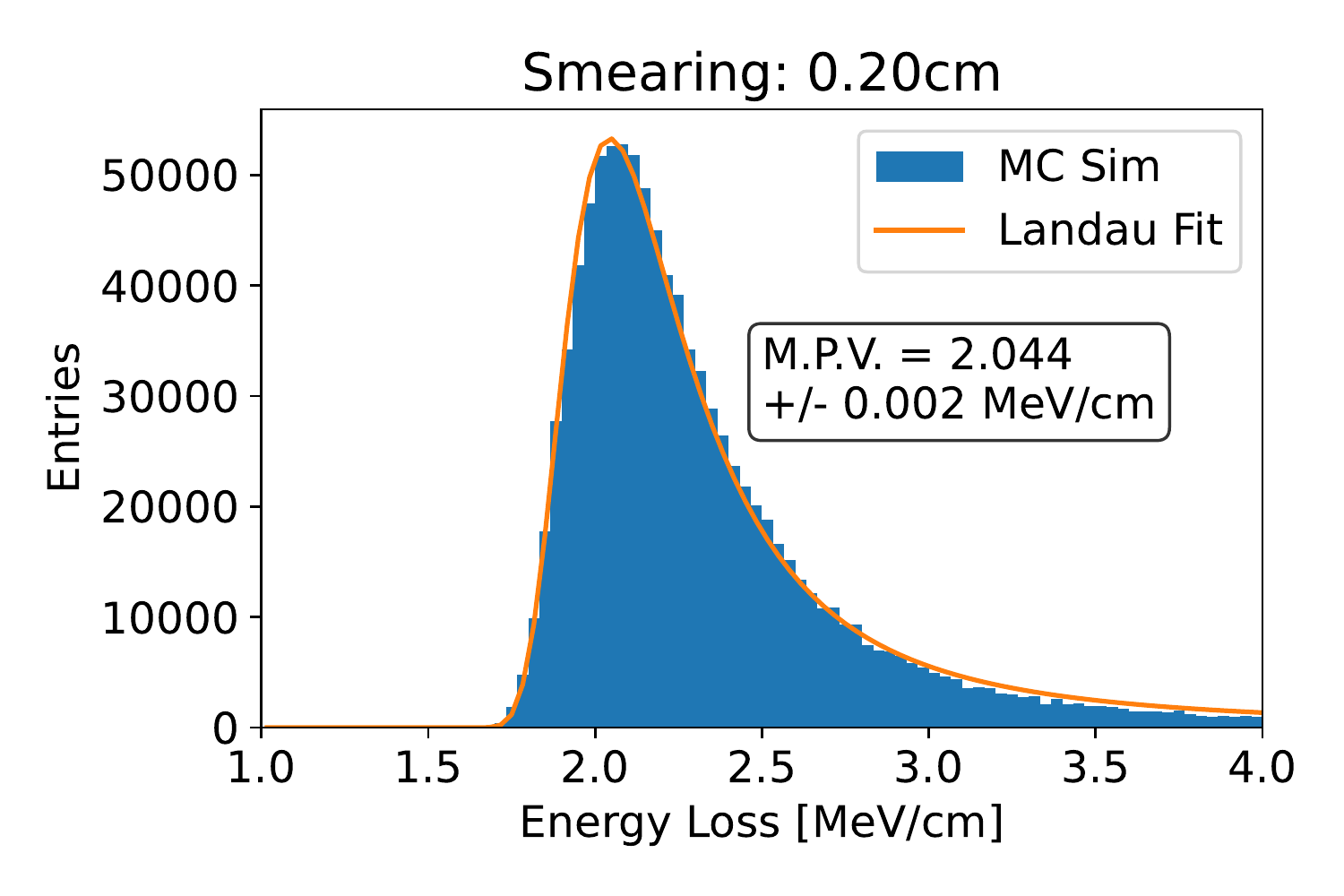}
	\includegraphics[width=0.22\textwidth]{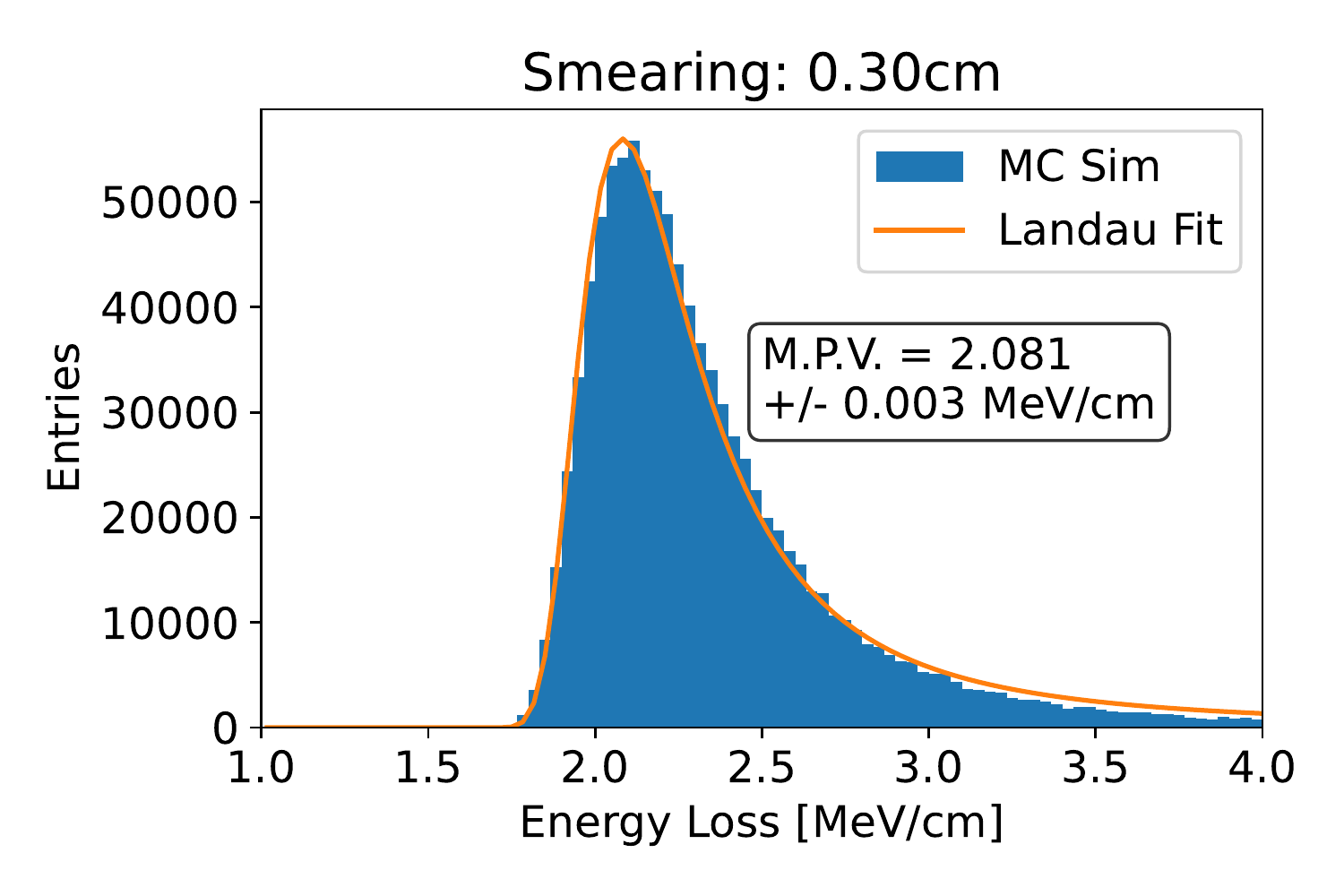}
	\hspace{1cm}
	\includegraphics[width=0.22\textwidth]{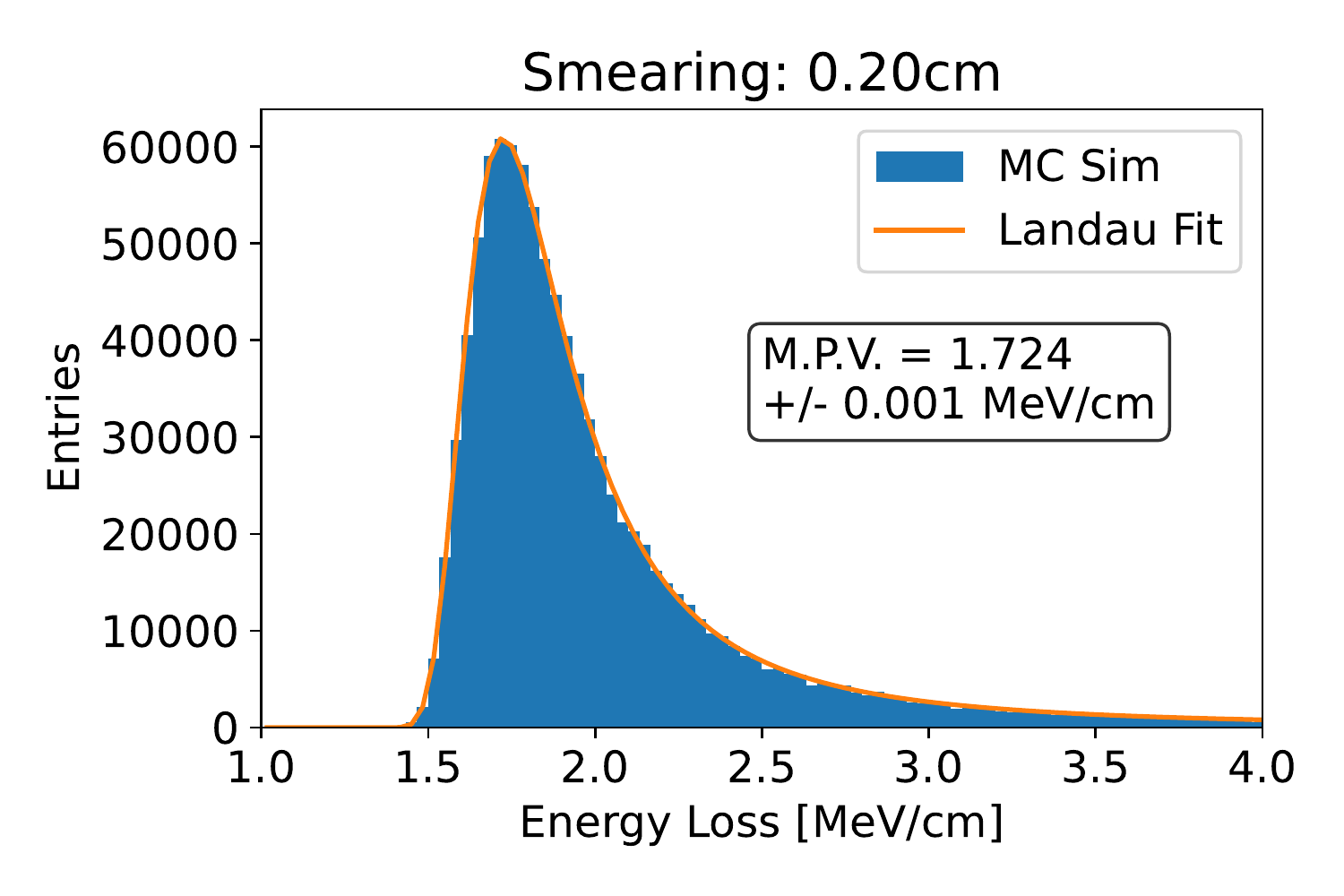}
	\includegraphics[width=0.22\textwidth]{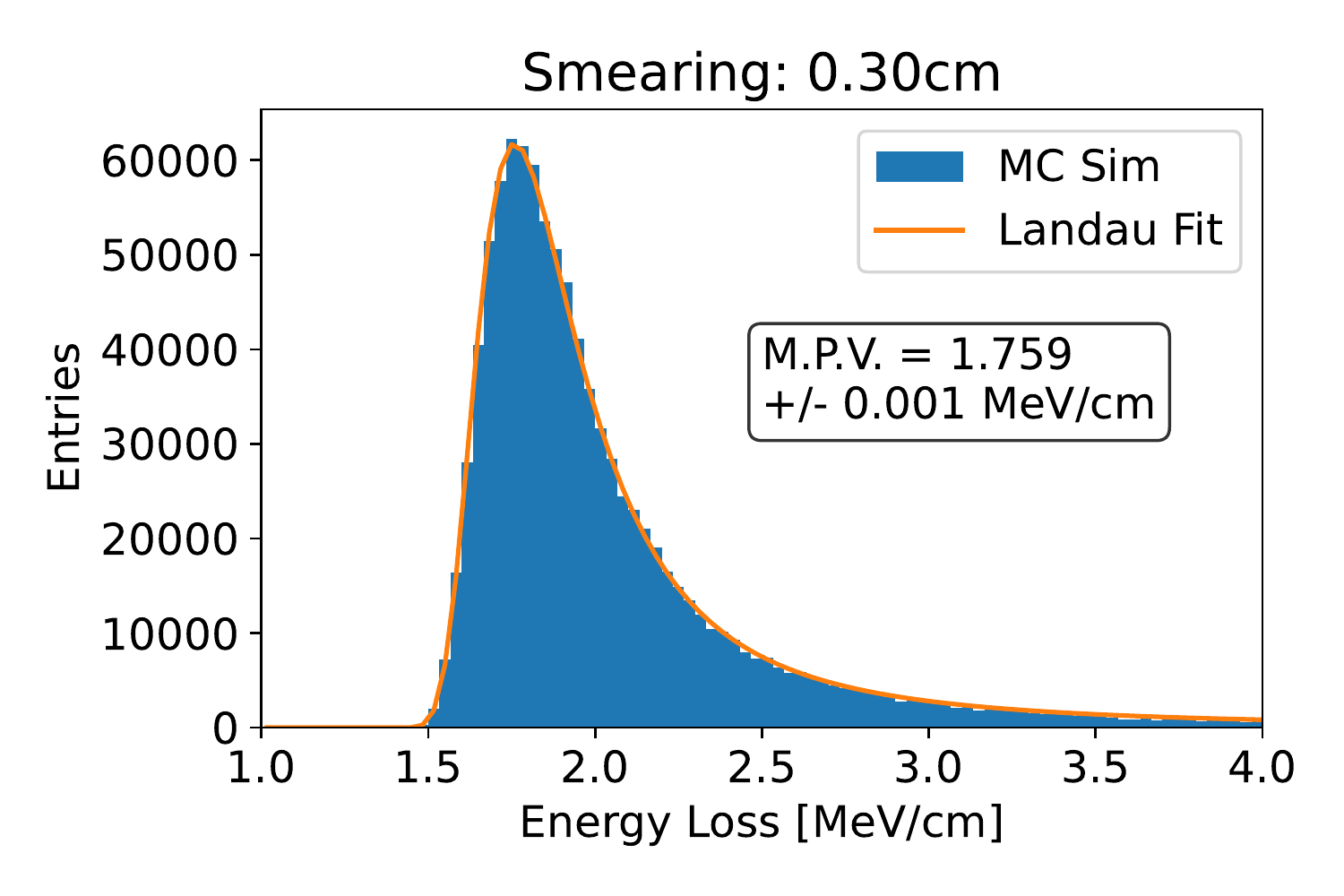}
	
	\vspace{0.5cm}
	
	Muon Energy: \SI{1}{GeV}
	\hspace{3cm}
	Muon Energy: \SI{10}{GeV}\\
	\includegraphics[width=0.22\textwidth]{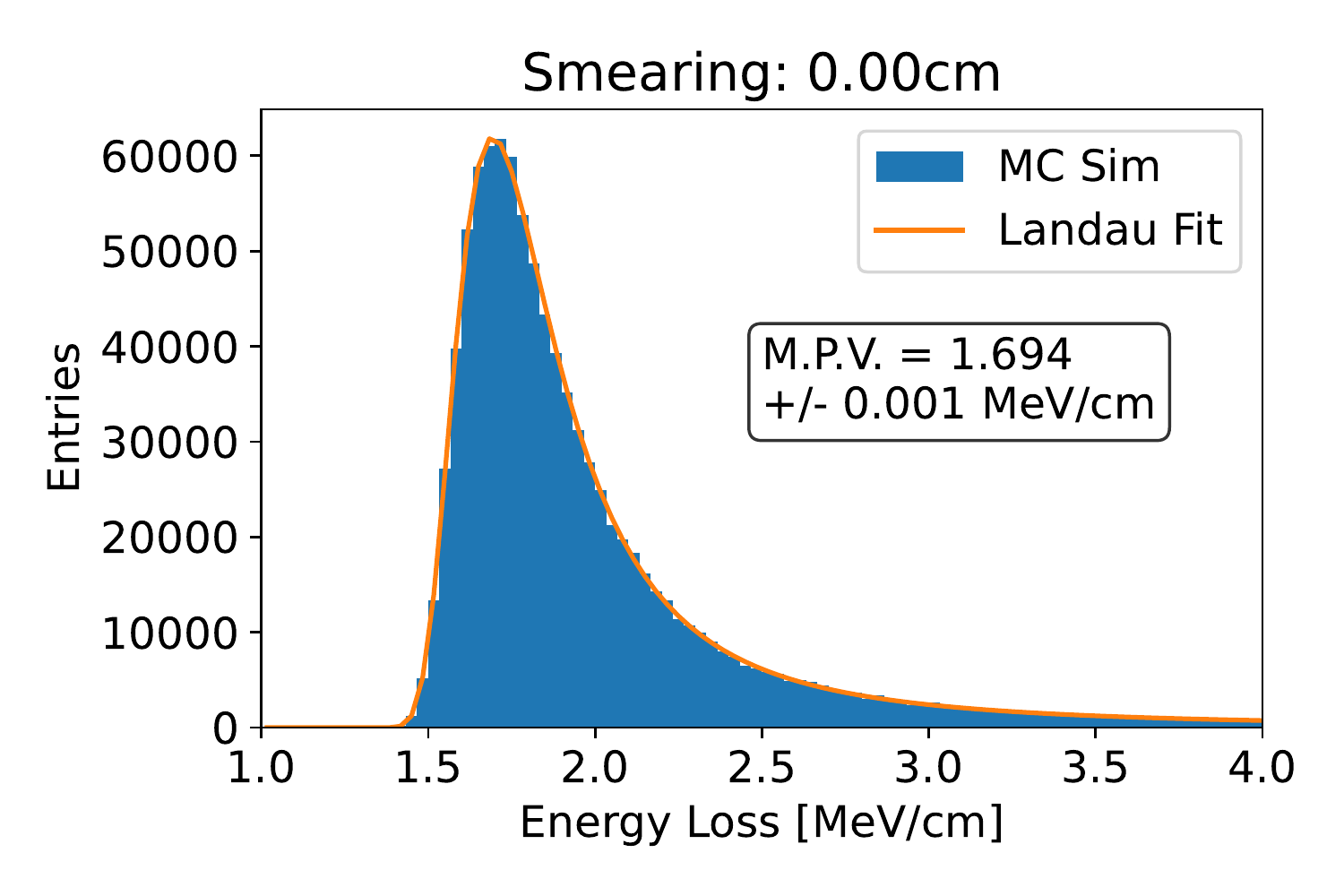}
	\includegraphics[width=0.22\textwidth]{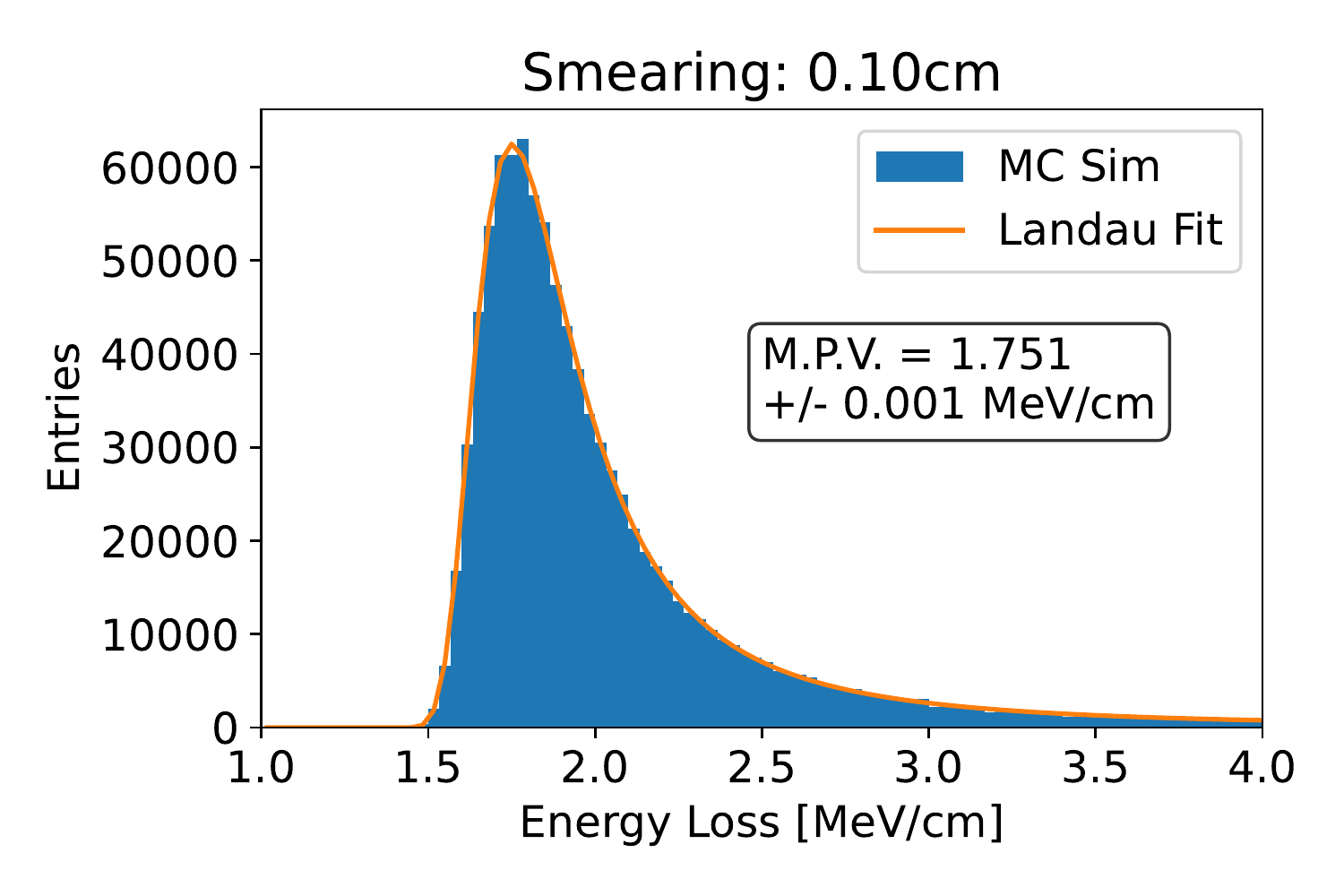}
	\hspace{1cm}
	\includegraphics[width=0.22\textwidth]{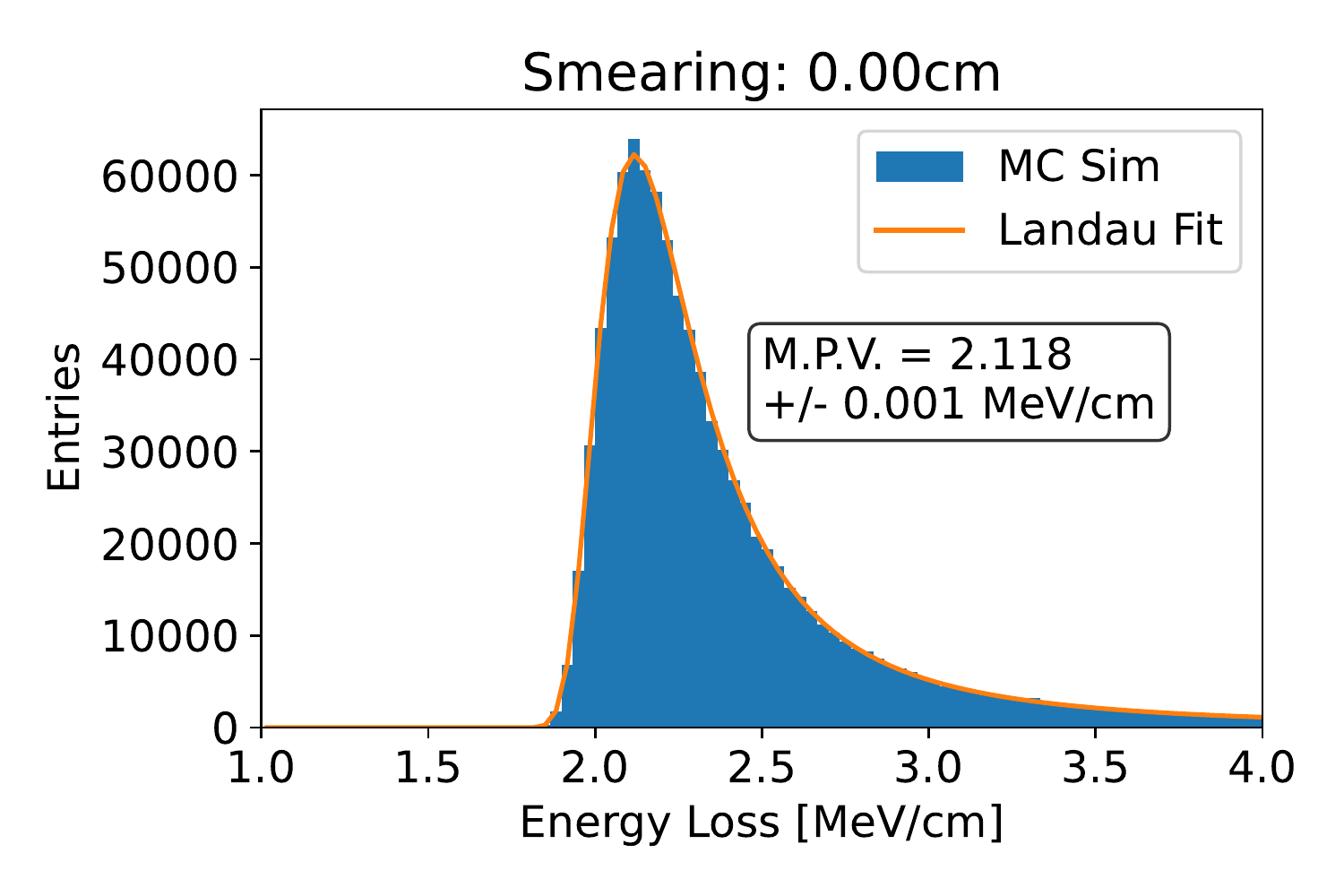}
	\includegraphics[width=0.22\textwidth]{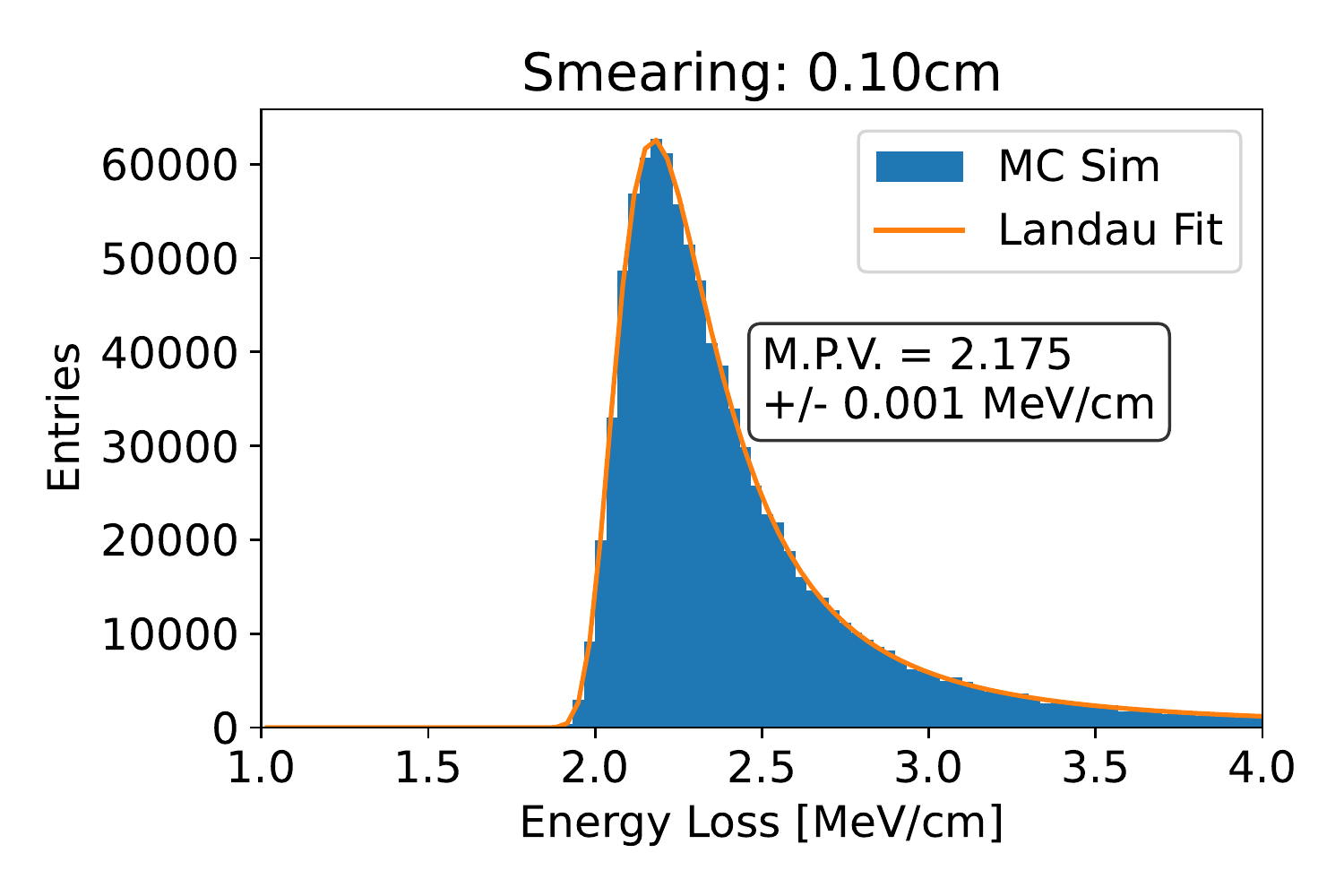}
	
	\includegraphics[width=0.22\textwidth]{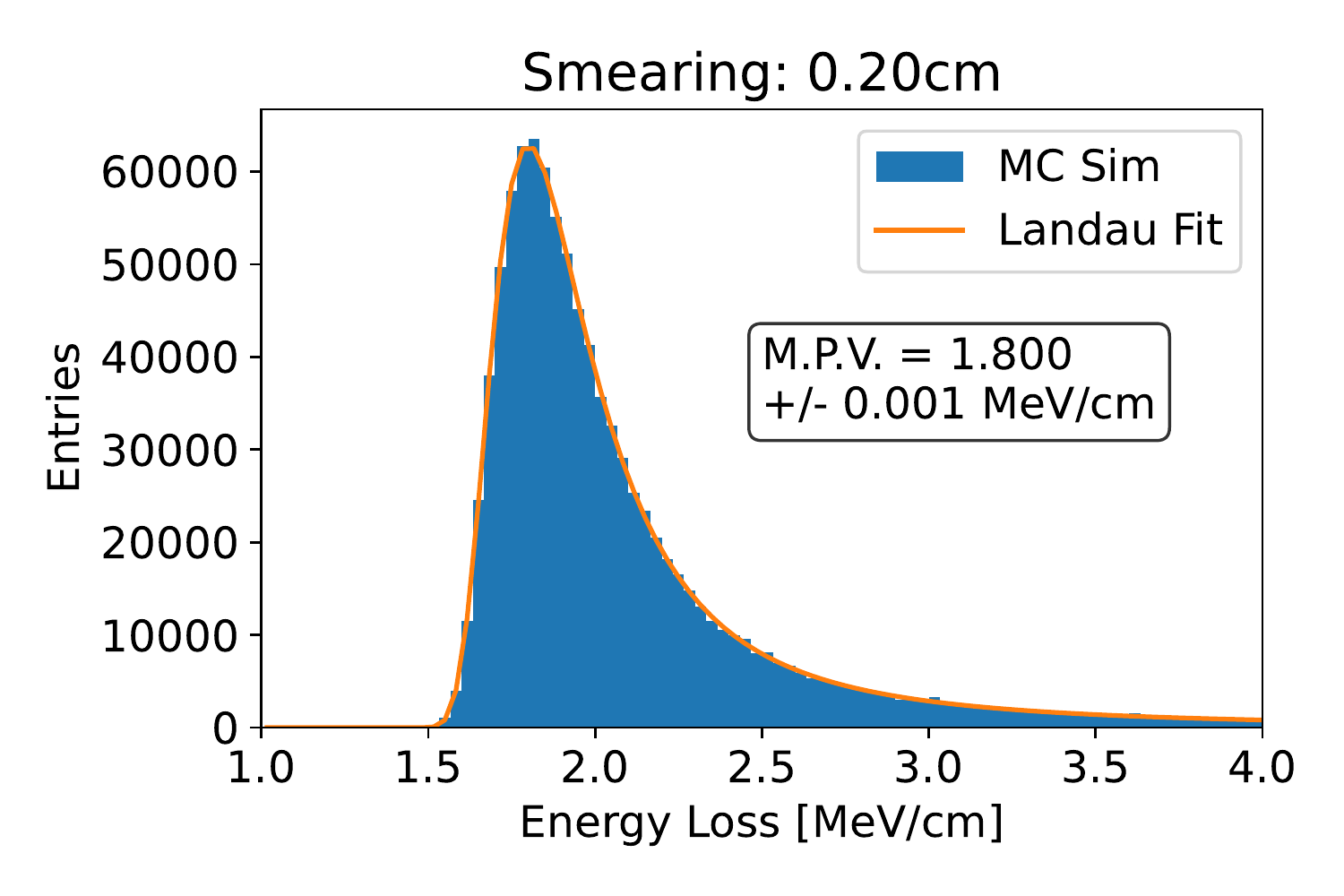}
	\includegraphics[width=0.22\textwidth]{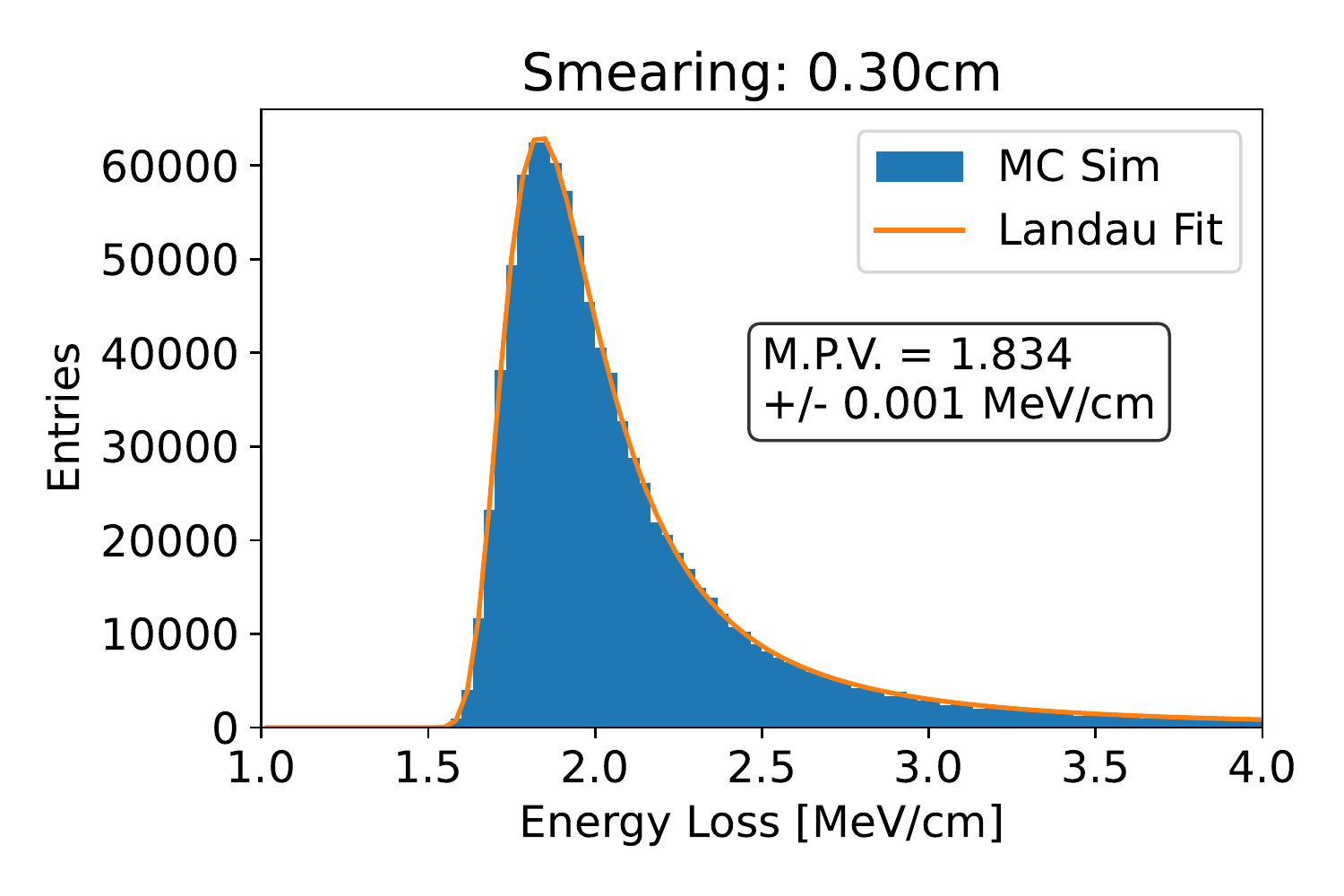}
	\hspace{1cm}
	\includegraphics[width=0.22\textwidth]{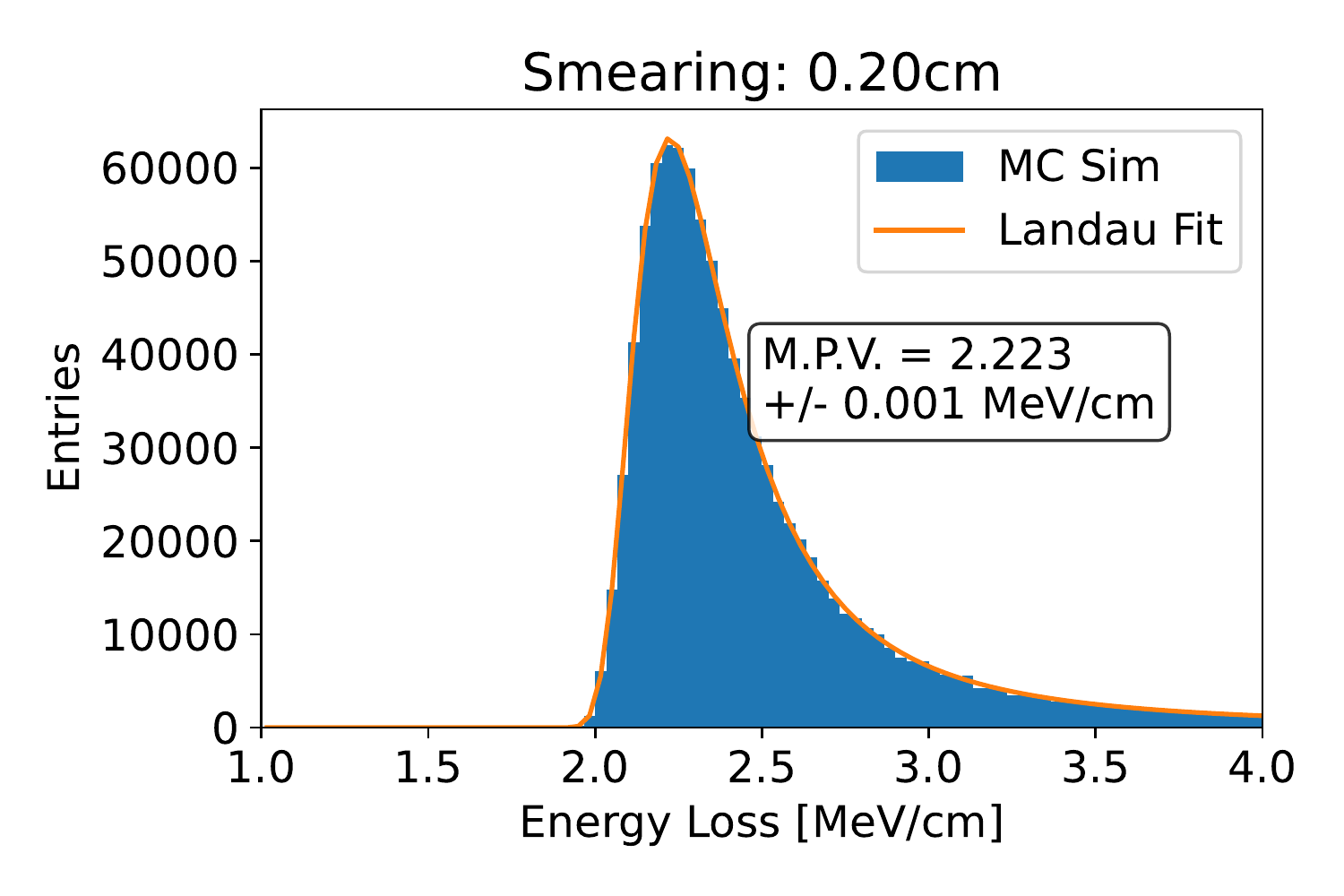}
	\includegraphics[width=0.22\textwidth]{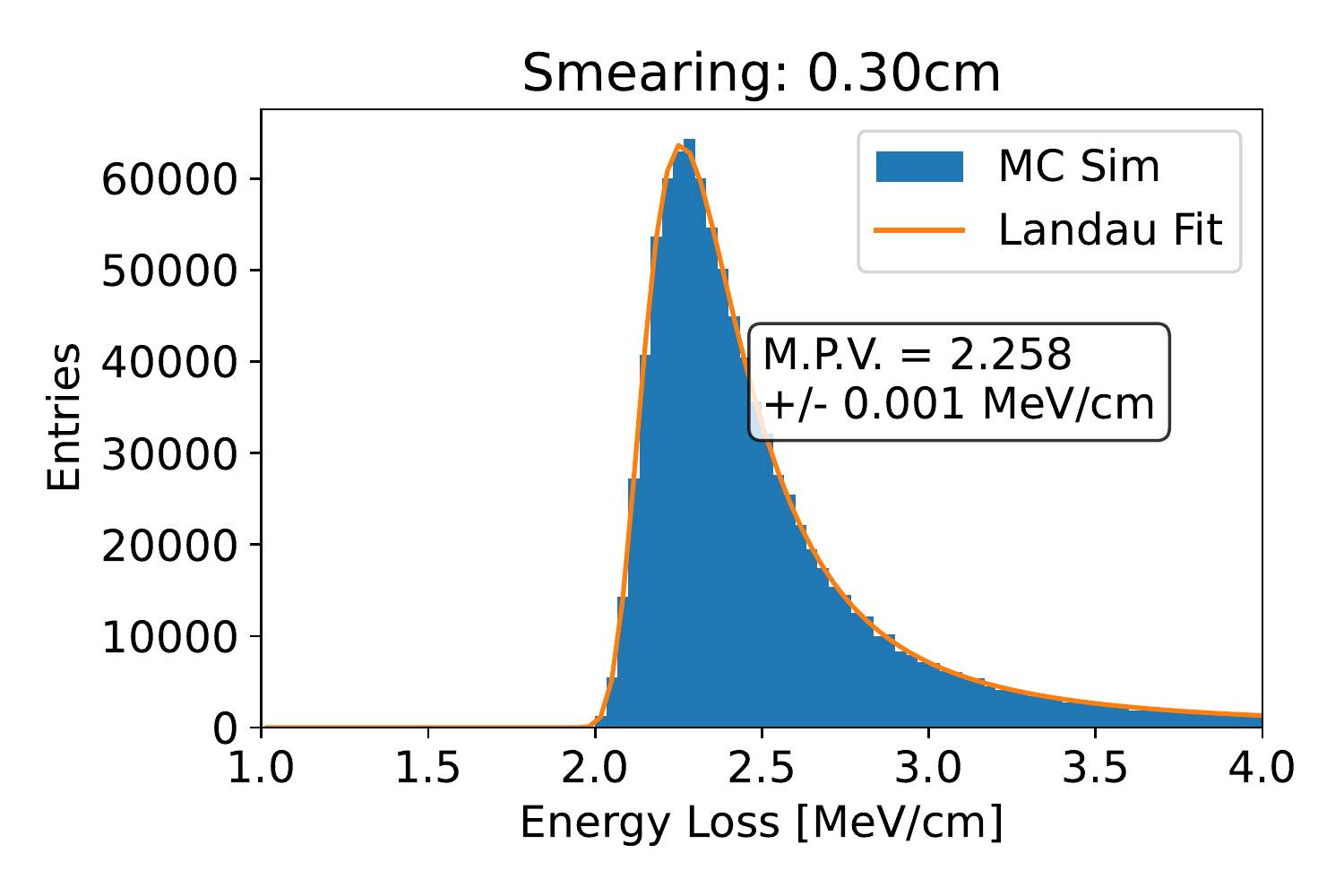}
	\caption{Distribution of energy loss values for various model parameters of the
	numerical computation. The fitted MPV of each run is shown.}
	\label{fig:edist}
\end{figure}

In each run, the simulation was performed 1 million times to provide a large distribution of muon
energy losses. Then, a Landau function was fit to the resulting distribution to obtain an MPV.
This fit was only done on the 20 bins on either side of the peak bin in order to avoid
difficulties in the fit trying to directly model the tail.
Statistical uncertainties are taken from the fit. Distributions for example model parameters are
shown in figure \ref{fig:edist}. Note that at small energy and large thickness (where the
thin-film approximation breaks down), the Landau distribution no longer provides a very good fit to the
distribution. This can be seen from the fact that the Landau distribution over-estimates the size of the
tail, especially for $E=$ \SI{0.2}{GeV}.
The extraction of the MPV from the Landau fit may not be as accurate in this region,
which is reflected by the larger uncertainty in the fit.

\begin{table}[htbp]
\centering
\rowcolors{2}{gray!25}{white}
\begin{tabular}{c | c }
	\rowcolor{gray!15}
	Quantity & Value (Units are MeV, cm, s, g) \\
	\hline
	\rowcolor{gray!15}
	\multicolumn{2}{c}{Fundamental Constants}\\
	Muon Mass ($M$)& 105.6\\
	Electron Mass ($m_e$) & 0.5110\\
	Classical Electron Radius ($r_e$) & 2.817940 $\cdot 10^{-13}$\\
	Avogadro Number & 6.0221409$\cdot 10^{23}$ \\
	\rowcolor{gray!15}
	\multicolumn{2}{c}{Material Properties (LAr)}\\
	Mean Excitation Energy ($I_0$) & 188.0 $\cdot 10^{-6}$\\
	Mass Density & 1.396 \\
	Mass Number & 39.9623 \\
	\rowcolor{gray!15}
	\multicolumn{2}{c}{MC Simulation Properties}\\
	Small-$\ell$ distribution length ($\epsilon$) & 0.01$\cdot 2^{-20} \approx $
	9.537$\cdot10^{-9}$\\ 
	Distribution step size ($\Delta x$) & 0.01\\
	Energy cutoff ($E_\text{max}$) & 10\\
	\#Points in distribution ($N_\text{MC}$) &  50$\cdot 10^6$ \\
	Simulated particle length ($\ell_\text{MC}$) & 10\\
	\#Simulations per data point & $10^6$\\
	Wire separation in weight function ($a$) & 0.3\\
\end{tabular}
\caption{Values of parameters in the numerical computation of muon energy loss measured by a
LArTPC-like channel. Where applicable, the symbol of the quantity used in the text is also
provided.}
\label{tbl:MCval}
\end{table}




\clearpage

\end{document}